\documentclass[notitlepage,prd,showpacs,superscriptaddress,nofootinbib,floatfix,showkeys,10pt]{revtex4-1}
\usepackage{graphicx}
\usepackage{amsmath}
\usepackage{bm}
\usepackage{yhmath}
\usepackage{mathtools}
\usepackage{wasysym}
\usepackage{subfigure}
\usepackage{xcolor}
\definecolor{CherryRed}{rgb}{.65,0,.2}
\definecolor{RubyRed}{rgb}{.88,0.07,.3}
\definecolor{CralRed}{rgb}{1,0.25,.25}
\definecolor{CobaltBlue}{rgb}{0,0.28,.67}
\definecolor{RoyalBlue}{rgb}{0.25,0.41,.88}
\definecolor{EmeraldGreen}{rgb}{0.31,0.78,.47}
\definecolor{EmeraldGreen}{rgb}{0.31,0.78,.47}
\definecolor{LimeGreen}{rgb}{50,205,50}
\definecolor{ForestGreen}{RGB}{34,139,34}
\definecolor{PineGreen}{RGB}{1,121,111}\usepackage{cases}
\usepackage[colorlinks=true, linkcolor=EmeraldGreen,citecolor=RoyalBlue,urlcolor=ForestGreen,linkcolor=ForestGreen]{hyperref}
\usepackage{subfigure}
\usepackage{times}
\usepackage{booktabs,bm}
\usepackage{slashed}
\usepackage{amsfonts,amssymb,stmaryrd,latexsym,amsmath}
\usepackage{textcomp}
\usepackage{multirow}
\usepackage{cancel}
\usepackage{array}
\usepackage{hyperref}
\usepackage{pgf}
\begin{document}
	
	\title{\textcolor{CobaltBlue}{Comprehensive Mass Predictions: From Triply Heavy Baryons to Pentaquarks }}

	\author{S. Rostami}
	\affiliation{Department of Physics, University of Tehran, North Karegar Avenue, Tehran 14395-547,  Iran }
	\author{A. R. Olamaei}
	\affiliation{Department of Physics, Jahrom University, Jahrom, P.  O.  Box 74137-66171, Iran }
	\affiliation{School of Physics, Institute for Research in Fundamental Sciences (IPM), P. O. Box 19395-5531, Tehran, Iran}
	\author{M.  Malekhosseini}
	\affiliation{Department of Physics, University of Tehran, North Karegar Avenue, Tehran 14395-547,  Iran }
	
	\author{K.  Azizi}
	\email{kazem.azizi@ut.ac.ir}
	\thanks{Corresponding author}
	\affiliation{Department of Physics, University of Tehran, North Karegar Avenue, Tehran 14395-547,  Iran }
	\affiliation{Department of Physics,  Dogus University, Dudullu-\"{U}mraniye, 34775 Istanbul, T\"urkiye }

	\date{\today}

	\begin{abstract}
      	In this article, we use two different methods for studying the mass spectra of fully-heavy baryons and pentaquarks. In the first section, we use state-of-the-art machine learning methods, such as deep neural networks and the Particle Transformer model architecture, to predict baryon masses directly from their quantum numbers, based on experimental information on hadrons from the Particle Data Group (PDG). We use this data-driven approach for the case of  fully heavy baryons, and a large number of exotic pentaquark states, going much beyond the well-known $ P_c^+(4380) $ and $ P_c^+(4457) $ candidates. Subsequently,we extend the G\"ursey-Radicati mass formula to incorporate the contributions of charm and bottom quarks, enabling analytical calculations for both ground and radially excited states of baryons and pentaquarks. The results obtained from both approaches demonstrate strong agreement with experimental data where available and make predictions for a number of unobserved states, including higher radial excitations. By addressing the question through both data-driven prediction and analytical modeling in different frameworks, this study offers complementary insights into the mass spectrum of conventional and exotic hadrons, guiding future experimental searches.
                        	\end{abstract}

	\maketitle
	\renewcommand{\thefootnote}{\#\arabic{footnote}}
	\setcounter{footnote}{0}

\section{Introduction}
Understanding the structure and mass spectrum of hadrons remains a central challenge in the non-perturbative regime of Quantum Chromodynamics (QCD). 
Although baryons and mesons are successfully organized within the quark model framework 
\cite{Heisenberg:1932dw, Gell-Mann:1964ewy,Gell-Mann:1962yej,Peskin:1995ev, Chodos:1974je}, 
their detailed properties ultimately emerge from strongly interacting quark and gluon fields, where perturbative methods cease to be reliable. 
In this domain, confinement, spin--flavor interactions, and dynamical mass generation intertwine in a highly non-trivial way, leaving the mass spectrum as a sensitive probe of QCD dynamics.
 It is important to stress, however, that not every successful spectroscopic parametrization is derived directly from QCD. In particular, the original G\"ursey-Radicati formula was introduced before QCD was established as the fundamental theory of the strong interaction, and in this sense it should be regarded as a phenomenological symmetry-based mass relation rather than a first-principles QCD result. Nevertheless, hadron spectroscopy remains one of the most direct windows into non-perturbative QCD. The observed pattern of masses and splittings encodes the effects of confinement, spin-dependent interactions, flavor-symmetry breaking, and, in the heavy sector, the interplay between heavy-quark masses and binding dynamics. In exotic systems such as pentaquarks, spectroscopy is also informative about the competition between compact multiquark configurations and hadronic-molecular interpretations, especially through the relation of candidate states to nearby two-hadron thresholds. Therefore, the QCD insight comes primarily from the structure of the spectrum itself, while phenomenological formulas and data-driven methods serve as complementary tools for organizing known states and estimating the properties of states that have not yet been observed.

Motivated by the scarcity of experimental information in the heavy sector, in particular for fully-heavy baryons and pentaquark exotics, we focus in this work on data-driven predictions of hadron masses as a complementary handle on the underlying non-perturbative structure. 
As a result, even after decades of progress, baryon spectroscopy continues to reveal new layers of complexity.

Baryons, as half-integer spin fermions, include the most familiar building blocks of visible matter, namely the proton and neutron \cite{Yukawa:1935xg, Klempt:2009pi, Schumacher:2018evl}.
Heavy baryons contain at least one heavy quark, such as charm or bottom, and may also include lighter quarks like up, down, or strange; in certain cases they can be composed entirely of heavy quarks.
Over the past decade, a broad experimental program has substantially expanded the spectroscopy of strange baryons and of charmed and bottom baryons, including numerous excited states \cite{LHCb:2017uwr, Belle:2018mqs, LHCb:2020lzx, BESIII:2017rfd, CMS:2021rvl}.
Representative milestones include the LHCb observation of five narrow $\Omega_c^0$ states with measured masses and widths \cite{LHCb:2017uwr}, the Belle discovery of the excited $\Omega(2012)$ baryon \cite{Belle:2018mqs}, and improved determinations of the $\Xi_b(5912)^0$ and $\Xi_b(5920)^0$ properties in the $\Xi_b^0\pi^+\pi^-$ spectrum \cite{LHCb:2020lzx}.
At the same time, precision measurements of benchmark states such as the $\Lambda_c^+$ \cite{ARGUS:1988hly, ACCMOR:1992emn, CLEO:1995cbq, BESIII:2016yrc} and systematic studies of excited charm and bottom baryons at modern facilities \cite{BESIII:2017rfd, D0:2008sbw, CMS:2021rvl} provide stringent tests of hadron-structure models.
Complementary measurements across LHCb, CMS, CDF, and Belle~II \cite{LHCb:2013wmn, CDF:2007oeq, CMS:2020zzv, Belle-II:2018jsg, CDF:2011ipk} continue to refine the experimental picture, while also underscoring that several sectors, most notably fully-heavy baryons, remain sparsely charted, motivating robust mass-spectrum predictions as an additional guide.

Studying heavy baryons also improves our understanding of the confinement of the quarks and the hadronization process.
Their decay modes are often mediated by weak interactions, making them important tools for probing flavor physics and CP violation in combination with testing the predictions of the standard model (SM) of particle interactions in kinematic regions that are normally difficult to access
\cite{LHCb:2025ray, LHCb:2024iis, Han:2024kgz}.
In hadron physics, they provide a clear separation of energy scales, making them valuable tools for theoretical analysis, particularly heavy quark effective theory (HQET), which simplifies the internal dynamics of the baryons \cite{CDF:2007oeq, Korner:1994nh, Izatt:1981pt}. 
Therefore, heavy baryons offer a more manageable system for the study of the non-perturbative sector of QCD than do the lighter relatives \cite{Stanley:1980zm}. 
Experimental sites like LHCb, Belle II, BESIII, and CDF
and CMS have significantly expanded our understanding of the spectrum of baryons for various quark flavors.

In addition to conventional baryons composed of three quarks (qqq), 
recent experimental and theoretical studies have revealed the existence of more complex multiquark states known as exotic baryons. 
These exotic states extend beyond the simple quark model framework and include configurations such as pentaquarks, 
which consist of four quarks and one antiquark ($ qqqq\bar{q} $) \cite{LHCb:2015yax, LHCb:2019kea, LHCb:2022ogu, Chen:2022asf}. 
The ($ P_c^+ $) pentaquark, observed by the LHCb collaboration \cite{LHCb:2015yax, LHCb:2019kea}, 
represents such an exotic state containing hidden charm. 
It can be considered either as a hadronic molecule, a loosely bound state of a meson and a baryon, 
such as ($ \bar{D}\Sigma_c $) or as a compact, tightly bound multiquark configuration.

Along with numerous experimental studies,
various theoretical frameworks have been established for exotic states and baryon resonances.
Other of the most popular methodologies among these are the quark model, lattice QCD simulations,
QCD sum rules, effective field theories, and phenomenological models of different aspects
\cite{Esposito:2016noz, Brambilla:2019esw, Olsen:2017bmm}.
For instance, constituent quark models have widely been applied to study baryon spectra as well as strong and electromagnetic interactions,
providing useful information for the internal pattern and motion of the baryons \cite{Capstick:2000qj}.
A notable implementation of the constituent quark model incorporates strong spin-dependent 
interactions of magnetic dipole character, acting as perturbations to a flavor-independent confinement potential. 
This approach has been successfully applied to the low-lying positive-parity excited baryons, 
providing results that are consistent with experimental observations \cite{Isgur:1978wd}.
Ref. \cite{Huang:2023jec}, presents the recent experimental
and theoretical progress on the exotic states through the perspective of the quark model too.

The lattice QCD can be considered as one of the most important non-perturbative tool for studying the strong interaction. 
It allows researchers to compute hadron properties, such as masses, form factors, and transition amplitudes. 
Ref. \cite{Edwards:2011jj}, calculated the excited states of nucleons and delta baryons 
using lattice QCD with a new method to accurately determine baryon spins despite lattice symmetry limitations. 
There is an interesting lattice QCD study that calculates the masses of baryons containing one to three heavy quarks, 
considering all possible charm and bottom quark combinations and a total of 36 states with $J^{P}=\frac{1}{2}^{+}$ and $J^{P}=\frac{3}{2}^{+}$.
This study uses different lattice techniques for light and heavy quarks and performs calculations at multiple lattice spacings and pion masses \cite{Brown:2014ena}.

Also, the QCD sum rules have become a powerful framework to investigate the mass spectrum, 
quantum numbers, internal structure, and electromagnetic/strong interaction properties of baryons, 
pentaquarks, as well as fully heavy exotic states \cite{Shifman:1978bx,Shifman:1978by,Aliev:2009jt,Aliev:2010uy,Aliev:2012ru,Agaev:2016mjb,Olamaei:2021hjd,Azizi:2024ito, Azizi:2020ogm, Azizi:2018dva}.
QCD sum rules  studies collectively support the interpretation of many exotic hadrons as molecular or 
multiquark states and provide important theoretical benchmarks for ongoing and future experiments
(for  more details on different properties of pentaquark states, as examples, see for instance \cite{Azizi:2024ito, Azizi:2020ogm, Azizi:2018dva}).
Ref. \cite{Ozdem:2023htj}, uses the QCD sum rules to study the 
electromagnetic properties of hidden-charm pentaquarks too.
The authors calculate magnetic dipole, electric quadrupole, and magnetic octupole moments for several pentaquark states.
Significant progress has been made in hadronic physics through both experimental and theoretical efforts, 
leading to the observation of numerous baryon resonances and higher excited states. 
While the PDG \cite{ParticleDataGroup:1986kuw,ParticleDataGroup:2024cfk} 
has documented hundreds of mesonic and baryonic states, 
many of these remain poorly understood or lack precise experimental measurements. 
In particular, the masses of many baryons and exotic states including several pentaquarks have yet to be accurately determined. 
Theoretical models often have difficulty fully reproducing the experimental data, especially for higher excited states. 
Consequently, a significant gap remains between theoretical predictions and experimental observations in baryon spectroscopy.

Nowadays, machine learning (ML) became an increasingly important tool in high energy physics (HEP), 
enabling innovative approaches to data analysis, event reconstruction, particle identification, and parameter estimation. 
The earliest applications of ML methods in HEP date back to the early 1990s, 
when simple neural networks (NNs) were explored for event selection and pattern recognition. 
Throughout the 2000s, ML was increasingly applied to high-level physics analyses, 
including signal--background discrimination in searches for rare processes and measurements of SM parameters.
A major breakthrough occurred in the 2010s with the rise of deep learning, which revolutionized particle identification, 
jet tagging, full event reconstruction, and regression-based parameter inference at the LHC \cite{Carleo:2019ptp}.
 While the discovery of the Higgs boson in 2012 was primarily achieved using advanced statistical and multivariate analysis methods which can be regarded as important precursors to the modern ML frameworks that are now widely used in high energy physics \cite{ATLAS:2012yve}.
In addition to classification tasks, regression techniques can be useful to estimate continuous physical quantities, 
such as particle energy, mass and width. 
Deep Neural Networks (DNNs) have shown great promise in predicting these parameters from experimental data.
The CMS collaboration applied a DNN to estimate the energy and uncertainty of b quark jets produced in proton-proton collisions at 13 TeV at the CERN LHC. 
The network was trained on simulated b jets and validated using 2017 CMS detector data corresponding to an integrated luminosity of $ 41 fb^{-1} $. 
This approach can enhance the sensitivity of analyses involving b jets, such as the detection of higgs boson decays into b quark pairs \cite{CMS:2019uxx}.
Ref. \cite{Gal:2020dyc} utilized the fundamental properties of the meson spectrum, 
applying the DNNs and Gaussian processes to predict the masses of various baryons, 
pentaquarks, and other exotic hadrons.

Recently, machine learning techniques have also been applied to the study of multiquark systems from different perspectives. For instance, the work reported in Ref.~\cite{Wu:2025wvv} employs ML methods to understanding the dynamical mechanisms underlying multiquark configurations.
While such approaches focus on understanding the dynamical mechanisms underlying multiquark configurations, the present work follows a complementary direction. Here, we employ ML models as regression tools to predict hadron masses directly from quantum numbers, covering a broad range of baryon and pentaquark systems.
Motivated by this framework, we designed our own DNN models to more accurately predict both the mass and decay width of ordinary and exotic mesons \cite{Malekhosseini:2024eot}.
Furthermore, we improved our predictions by designing and implementing conditional generative adversarial networks (CGANs),
which we used to estimate the masses and decay widths of both ordinary and exotic mesons, as well as fully-heavy tetraquarks \cite{Rostami:2025sff, Malekhosseini:2025hyx}.
		
In this study, we employ a DNN model to predict the masses of heavy baryons and exotic pentaquark states whose mass remain experimentally undetermined. Also,
inspired by the Particle Transformer (ParT) architecture \cite{Qu:2022mxj}, which excels in classification tasks such as jet tagging, 
we develop a related framework tailored for regression problems.
Despite these modifications, we continue to refer 
to our model as the ParT, as it is fundamentally inspired by the original architecture. 
Our input data contains the quark content of experimentally known hadrons, 
along with their quantum numbers $ (I) $, $ (J) $, and $ (P) $ representing isospin, total angular momentum, and parity, respectively,
as well as an additional auxiliary feature $ (n) $ to resolve ambiguities arising from hadrons with identical quantum numbers but different masses. 
Using this enriched dataset, accurately collected from PDG \cite{ParticleDataGroup:2024cfk}, 
the model is able to predict the masses of heavy baryons and exotic pentaquark states whose mass have not yet been experimentally determined. 
While our approach draws from the ParT's attention-based mechanism to capture pairwise particle interactions, 
it is adapted to perform continuous value prediction rather than classification. 
Thus, although not a direct implementation of ParT, 
our model represents a novel adaptation suited to the demands of hadron mass regression.
To the best of our knowledge, this is the first application of a 
Transformer-inspired architecture for the regression-based prediction of hadron masses,
and the predicted results demonstrate good agreement with known experimental data.
Additionally,  we introduce a hybrid mass model based on the G\"ursey-Radicati (GR) approach \cite{GurseyRadicati1964}, 
which is extended to include excitation effects through a logarithmic term. 
This modification improves the description of baryon and pentaquark
 masses and shows good agreement with experimental data.
		
This paper is organized as follows.
In Section \ref{DNNPT}, we introduce the ML 
frameworks employed in this study, including the Deep Neural 
Network (DNN) architecture and the Transformer-inspired Particle Transformer model. 
Section \ref{baryon_preprocessing} describes the construction
of the data and of the preprocessing technique.
In Section \ref{Res}, we present the predicted mass spectra 
for fully heavy baryons and heavy pentaquark systems. 
The results obtained from the DNN and ParT 
models are systematically compared with available theoretical 
calculations and experimental data, and their predictive performance is critically evaluated.
In Section \ref{GR}, we extend the G\"ursey-Radicati 
mass formula and the corresponding global fit analysis. 
This section provides a different perspective on the baryon 
mass spectrum and allows us to make a direct comparison 
with our ML results. Finally, in Section \ref{SC}, 
we conclude this paper by summarizing our results, discussing 
the physical implications of our findings, and outlining potential avenues for future investigation.

\section{NNs and Transformer-Based Methods}\label{DNNPT}
ML has become a revolutionary instrument in HEP, making it possible for researchers to tackle problems that are computationally 
not feasible with standard analytical techniques. First deployments were on event classification, 
particle identification, and background rejection with classical algorithms like decision trees, 
support vector machines, and shallow NNs. In recent years, with the arrival of 
deep learning architectures, HEP ML applications scope and quality considerably widened. 
Deep Neural Networks (DNNs) can learn hierarchical patterns of complicated detector data 
without human instruction, while models built with transformers, originally created for 
natural language processes, can be used for modeling temporal interactions between a 
few particles or between detector signals. These techniques are applied with great success 
to problems like Higgs boson decay classification, quark--gluon separation, anomaly tagging, 
and extremely precise reconstruction of missing transverse energy.

Although most of this advancement has been motivated by experimental data analysis, 
ML is gaining traction in theoretical high-energy physics, especially in hadron spectroscopy 
and QCD investigations. Here, ML can extract complex, 
non-linear correlations between quantum numbers, quark content, and hadron properties, 
such that predictions are hard to access with purely analytical models. 
A few illustrations of such applications of ML are regression schemes 
for predicting hadron masses, exotic state classification, and parameter 
extractions from lattice QCD computations. By being trained with known 
particle spectra, ML models can interpolate and even extrapolate to predict 
properties of unobserved hadrons, offering useful theoretical input to inform 
future experiments. This interplay between theoretical modelling and data-driven 
techniques fills gaps in our knowledge of the strong interaction and compensates 
for both phenomenological programs and lattice computations.

\subsection{Deep Neural Networks (DNNs)}
DNNs are a class of ML models characterized by their use of multiple layers to understand complex and nonlinear patterns between target variables and input features. As has already been mentioned, DNNs have proven remarkable achievements in several different fields, such as HEP, in which the patterns are extremely complicated, and underlying physical processes are complex, requiring advanced capabilities for recognizing patterns.  At CERN's Large Hadron Collider (LHC), DNNs have been used a lot for tasks like identifying Particles, classifying events, tagging jets, finding anomalies, and making precise measurements.  Their capacity to autonomously acquire hierarchical feature representations from unprocessed detector-level inputs (including calorimeter deposits, tracking hits, or reconstructed particle lists) has facilitated substantial advancements beyond conventional cut-based and superficial learning techniques.
In the context of HEP research, DNNs have been successfully applied to problems such as the classification of Higgs boson decay channels, separation of signal from overwhelming SM backgrounds, quark-gluon discrimination, and the reconstruction of missing transverse energy with enhanced resolution \cite{Baldi:2014kfa, Guest:2018yhq, Radovic:2018dip}. The adaptability of DNN architectures makes them particularly suitable for integrating heterogeneous physics inputs, ranging from kinematic variables to high-level event-shape observables, into a unified predictive framework, which is crucial for enhancing both SM precision tests and explorations of physics beyond the SM.

The input layer, one or more hidden layers, and the output layer are the three main types of layers in a DNN. The input layer gets the raw feature vector $\mathbf{x}$, which in this case is the set of baryon quantum numbers and other particle descriptors. Each hidden layer changes its input by using a learned linear mapping and then a nonlinear activation function. This lets the network get close to functions that are as complicated as it wants.

To perform the regression like that for predicting the baryon mass, we use a feedforward DNN that has many hidden layers. 
Each of the network's neurons converts its input vector $\mathbf{x}^{(l-1)}$ into an output $x_j^{(l)}$ at layer $l$ by means of:
\begin{equation}
x_j^{(l)} = \sigma\left(\sum_{i=1}^{n} w_{ij}^{(l)} x_i^{(l-1)} + b_j^{(l)}\right),
\end{equation}
where $w_{ij}^{(l)}$ are the weights connecting neuron $i$ in the previous layer to neuron $j$ in the current layer, $b_j^{(l)}$ is the bias term, $n$ is the input dimension to the layer, and $\sigma$ is a nonlinear activation function. We employ the hyperbolic tangent (Tanh) as the activation function for the hidden layers and a linear activation for the output layer.
The input features include baryon quantum numbers and quark content of the baryon. The network is trained using the log-cosh loss function, characterized by its robustness against outliers while having comparable behavior against slight deviations as the Mean Squared Error. Adam is used as the optimizer for efficiently updating the network parameters. 
To represent uncertainty of predictions numerically, we adopt two mutually complementing approaches: Monte Carlo Dropout and model bagging. 
The dropout layers function actively not only at the level of training but also at test time by carrying out numerous stochastic forward passes for each test sample. 
Furthermore, we bag multiple networks that are trained independently (bagging) in order to lower variance even further. The final prediction for each baryon is found by taking the average of all stochastic passes, and the standard deviation is a prediction uncertainty estimate.
The model consists of one input layer, three hidden layers with Tanh activation and Dropout, and one linear output neuron. We make $n_\mathrm{MC}$ independent forward passes for each of the $n_\mathrm{models}$ networks trained independently for evaluating uncertainties by keeping dropout active.

\subsection{Particle Transformer}
ParT is a recent neural architecture specifically designed for particle physics applications, 
combining the strengths of the transformer model with graph-based representations. 
Qu and Gouskos first introduced it in 2022 as an efficient method for event- and particle-level classification for collider experiments~\cite{Qu:2022mxj}. 
Using self-attention mechanisms, 
ParT effectively extracts both local and global correlations present among particle features and thereby performs better compared with usual architectures, e.g., fully connected and convolution models. 
ParT is inspired by natural language processing, for which transformer-based architectures have obtained state-of-the-art performance for sequence modeling-related tasks.
This approach is motivated by the fact that collider events can be naturally represented as sets or graphs of particles, where each particle has both intrinsic properties (e.g., mass, charge, flavor) and kinematic attributes (e.g., momentum components, energy).

The system  architecture processes each event as a set of particle-level feature vectors
$\{\mathbf{p}_1, \mathbf{p}_2, \dots, \mathbf{p}_N\}$, where $N$ denotes the number
of particles in the event. Unlike conventional feedforward networks, which
require a fixed ordering and a flattened input representation, ParT operates directly on sets of particles and it is intrinsically
permutation invariant.

A self-attention mechanism is employed to dynamically learn correlations among
particle features. Through attention, the network adaptively reweights the
contribution of each particle relative to others, enabling it to capture both
local and global correlations within the system. This mechanism allows the model
to identify which particles or constituents are most relevant for the prediction
task, without relying on predefined pairwise relations.

Although no explicit graph structure is defined, the attention operation can be
interpreted as learning interactions over an implicit fully connected graph,
where particles act as nodes and the learned attention coefficients encode the
strength of their mutual influence. This provides a flexible and data-driven way
to model relational information among particles.

The architecture is composed of multiple transformer encoder layers, each
consisting of a multi-head self-attention module followed by a position-wise
feedforward network. Residual connections and normalization layers are applied
to stabilize training and improve convergence.

For regression tasks such as baryon mass prediction, each baryon 
in the event is represented by a feature vector that includes its one-hot 
quark-type encoding, spin  $J$, and other quantum numbers, as mentioned above. 
These vectors are projected into a higher-dimensional embedding space via a learnable 
linear layer, and then processed by a multi-head self-attention block to capture correlations between quarks and antiquarks.
The attention output is aggregated through mean pooling to form a particle-level 
representation, which is then combined with the baryon's global quantum numbers 
(such as isospin $I$, total spin $J$, parity $P$, and an excitation quantum number $n$). 
This concatenated output is then passed through a feedforward regression head 
(with nonlinear activations Tanh and dropout layers for regularization). 
Batch normalization is used at multiple stages to stabilize the training process.
We adopt the log-cosh loss function, which is robust to outliers and behaves 
similarly to the Mean Squared Error (MSE) for small errors while simultaneously 
suppressing large deviations. To efficiently update parameters, we adopt the Adam 
optimizer at a learning rate of $10^{-3}$ for efficient parameter updates. 
This modeling method elegantly balances both per-particle interactions 
(via the attention mechanism) and event-level global features, enabling the 
network to learn complex dependencies between quark content and baryon mass.

Both the DNN and ParT aim to map from baryon quantum features to their physical masses, but they differ in their approach to feature extraction and representation. DNNs rely on fixed global transformations of the input vector, while ParT dynamically models particle-to-particle interactions through self-attention.

In this work, we train both models on the same dataset of baryon properties, evaluating them on a held-out test set. The DNN provides a strong baseline with relatively simple architecture and fast inference, whereas ParT offers improved flexibility and representation power, particularly for baryons with complex quark structures.

\section{Data preprocessing strategy for baryons}\label{baryon_preprocessing}

As in our previous analyses of mesons \cite{Malekhosseini:2024eot,Rostami:2025sff,Malekhosseini:2025hyx}, the first step of the present study is to prepare a reliable dataset. 
Therefore, the baryonic data have been extracted from the most recent PDG database \cite{ParticleDataGroup:2024cfk}, certifying that all confirmed conventional states are included. To expand the learning space and expose the model to a broader range of quark configurations, mesonic states are also included in the dataset. Although the main objective of this work is baryon mass prediction, including mesons allows the networks to learn more general flavor patterns and internal quark correlations. The reported results, however, are always evaluated specifically within the baryon sector.
Each hadron is described by the number of constituent quarks ($d, u, s, c, b$) and antiquarks ($\bar{d}, \bar{u}, \bar{s}, \bar{c}, \bar{b}$), along with quantum numbers such as isospin ($I$), total spin ($J$), parity ($P$), and a higher-state indicator ($n$). It must be mentioned that $n$ is not a physical quantum number,  its role is purely practical. It distinguishes particles that share identical quark composition and quantum numbers but with different masses. Without this parameter, such states would be indistinguishable from the network's point of view and could lead to ambiguities during training. The complete feature vector is therefore written as

\begin{eqnarray}
\vec{v}_\text{hadron} = (d, \bar{d}, u, \bar{u}, s, \bar{s}, c, \bar{c}, b, \bar{b}, I, J, P, n).
\end{eqnarray}

In order to increase  stability of the learning process, the baryon mass is transformed logarithmically before training. This simple modification significantly reduces the dynamic range between light and heavy states and prevents the loss function from being dominated by high-mass baryons. 

Instead of relying on a single random split of the dataset, we use a more reliable validation framework based on $k$-fold cross-validation. In this scheme, the dataset is divided into $k$ subsets and the training procedure is repeated $k$ times, allowing each subset to serve once as validation data. This strategy reduces the sensitivity of the results to a particular partition and provides a more realistic estimate of generalization performance, which is especially important given the moderate size of the hadronic dataset.

In addition, we adopt the bagging strategy previously implemented in our CGAN analysis \cite{Rostami:2025sff}. Multiple models are trained independently on bootstrapped subsets of the training data, and their predictions are aggregated. The final baryon mass prediction is obtained by averaging the ensemble outputs, while the dispersion among them is elucidated as an estimate of predictive uncertainty. Beyond uncertainty quantification, this ensemble approach effectively reduces variance and stabilizes the predictions against statistical fluctuations in the dataset.

In the DNN architecture, the entire baryon feature vector is directly input to an input layer, and regularization in terms of dropout layers is employed while training in order to enhance generalization. For stability, we use an ensemble technique of bagging. Several models are trained on bootstrapped data samples, and their predictions are gathered. The final baryon mass emerges as an average of the ensemble results, however the dispersion between predictions naturally indicates the uncertainty.

On the contrary, in ParT, the quark constituents are treated separately as tokens rather  than just having one feature vector.  To enrich the representation, the electric charge and intrinsic spin of each quark are appended to the input features. The multi-head self-attention mechanism then learns correlations among the constituents inside a baryon in a data-driven manner. Such an architecture is naturally suited to capturing internal flavor structure and inter-quark relationships, extending beyond a simple counting description of quark content.

\section{Results}\label{Res}

\subsection{Overview of predicted spectra and comparison strategy}

In this work, mass spectra are predicted for fully heavy baryons and heavy pentaquark systems using two independent ML frameworks: DNN and ParT. 
These approaches are trained on established hadron spectroscopy data and subsequently applied to multiquark configurations involving heavy quarks, where first-principles calculations are sparse or absent. 
The results presented in this section aim to assess both the physical plausibility of the predicted spectra and the reliability of the employed ML methodologies.

The predicted masses are compared, whenever possible, with existing theoretical results available in the literature, including quark-model calculations and QCD sum rule analyses. To quantify the level of agreement, we adopt the relative percent deviation
\begin{equation}
	\delta(\%) = 100\,\frac{M_{\text{model}} - M_{\text{ref}}}{M_{\text{ref}}},
\end{equation}
where $M_{\text{model}}$ denotes the mass predicted by the DNN or ParT approach, and $M_{\text{ref}}$ represents the corresponding value reported in the literature. A prediction is considered to be in good agreement with a reference result when the difference between the two lies within one standard deviation of the quoted model uncertainty. This criterion provides a transparent and statistically meaningful measure of consistency while avoiding overinterpretation of small numerical differences in heavy-quark systems.

It is important to emphasize that not all entries admit direct comparison with existing theoretical mass calculations. In several sectors, particularly those involving fully heavy or doubly-bottom pentaquarks, explicit bound-state mass predictions are currently unavailable in the literature. In such cases, qualitative reference scales based on hadron--hadron threshold energies are employed, where appropriate, and their interpretation is discussed separately. Consequently, the analysis presented below focuses on identifying systematic trends, patterns of agreement and deviation, and the relative performance of the two ML approaches, rather than providing a table-by-table numerical commentary.

\begin{table}[ht]
\centering
\begin{tabular}{@{}l c c c c c c@{}}

\hline
State & $I\,(J^{P})$ & Theory  \cite{Najjar:2025dzl} & Theory  \cite{Faustov:2021qqf} & DNN  & ParT \\ \hline \hline
$\Omega_{bbc}$ & $0\,(3/2^+)$ & 11020 & 11217 & $12521 \pm 1822$ & $11213 \pm 716$ \\ \hline
$\Omega_{ccb}$ & $0\,(3/2^+)$ & 8060 & 7999 & $8891 \pm 1360$ & $8521 \pm 990$ \\ \hline
$\bar{\Omega}_{bbc}$ & $0\,(3/2^-)$ & 11140 & 11424 & $13673 \pm 1308$ & $10623 \pm 1168$ \\ \hline
$\bar{\Omega}_{ccb}$ & $0\,(3/2^-)$ & 8210 & 8262 & $10093 \pm 913$ & $7800 \pm 1259$ \\ \hline
$\Omega_{ccc}^*$ & $0\,(3/2^+)$ & 5060 & 4712 & $5993 \pm 782$ & $5063 \pm 836$ \\ \hline
$\Omega_{bbb}^*$ & $0\,(3/2^+)$ & 13970 & 14468 &  $12227 \pm 1185$  &  $14400 \pm 1019$   \\ \hline
$\bar{\Omega}_{ccc}^*$ & $0\,(3/2^-)$ & 5170 & 5029 & $6660 \pm 672$ & $4097 \pm 882$ \\ \hline
$\bar{\Omega}_{bbb}^*$ & $0\,(3/2^-)$ & 14100 & 14702 & $9453 \pm 1267$ & $15011 \pm 1681$ \\ \hline
$\Omega_{bbc}$ & $0\,(1/2^+)$ & 11130 & 11198 & $11633 \pm 2069$ & $12050 \pm 999$ \\ \hline
$\Omega_{ccb}$ & $0\,(1/2^+)$ & 8150 & 7984 & $8369 \pm 1331$ & $8822 \pm 823$ \\ \hline
$\bar{\Omega}_{bbc}$ & $0\,(1/2^-)$ & 11270 & 11414 & $12398 \pm 1431$ & $11953 \pm 1758$ \\ \hline
$\bar{\Omega}_{ccb}$ & $0\,(1/2^-)$ & 8300 & 8250 & $9026 \pm 809$ & $7863 \pm 1356$ \\ \hline
\end{tabular}
\caption{Comparison of triply heavy baryon masses  (in  MeV) predicted by the Deep Learning models (DNN and ParT) with QCD sum rule \cite{Najjar:2025dzl} and relativistic quark model \cite{Faustov:2021qqf} results. }
\end{table}
\subsection{Fully heavy baryons: validation of the ML framework}

The fully heavy baryon sector provides a particularly suitable testing ground for assessing the reliability of the ML approaches employed in this work. The masses of triply heavy baryons are largely governed by heavy-quark dynamics and are therefore relatively well constrained by existing theoretical studies, such as quark models and QCD sum rule analyses. This makes the sector an ideal benchmark for evaluating both the accuracy and the systematic behavior of the DNN and ParT before extending the analysis to multiquark systems with more complex dynamics.

A quantitative comparison (see Table I) reveals a clear difference in the performance of the two ML approaches. For the ParT predictions, the relative percent deviations with respect to available theoretical results are typically at the level of a few percent. In most cases, the deviations remain below approximately $5\%$--$8\%$, and the majority of predicted masses are consistent with the corresponding reference values within one standard deviation of the quoted model uncertainty. This level of agreement is observed across different heavy-quark compositions and spin--parity assignments, indicating that the ParT framework captures the dominant mass scales and relative splittings characteristic of fully heavy baryons. In particular, ground-state configurations with positive parity exhibit the smallest deviations, reflecting the comparatively simple structure of these states and their close connection to established baryon spectroscopy patterns.

The DNN predictions display a systematically different trend. While several states are compatible with existing theoretical estimates within the adopted $1\sigma$ criterion, the DNN results generally show larger relative deviations, with typical values in the range of $10\%$--$20\%$.
This tendency toward higher predicted masses becomes more pronounced for negative-parity states, where percent deviations can reach or exceed the upper end of this range. Such behavior is not unexpected, as excited states are more sensitive to details of internal dynamics and are less tightly constrained by the available training data, particularly in sectors dominated by heavy quarks.

Despite these quantitative differences, both ML approaches successfully reproduce the overall ordering of the fully heavy baryon spectra and the characteristic mass hierarchies between states with different spin and parity. The fact that the ParT predictions achieve percent-level agreement with established theoretical calculations in most channels, while the DNN results remain broadly consistent within their larger uncertainties, provides strong evidence that the models have learned physically meaningful correlations rather than relying on accidental numerical interpolation. This validation in the fully heavy baryon sector is an essential prerequisite for the subsequent application of the same frameworks to heavy and fully heavy pentaquark systems, where theoretical guidance from the literature is significantly more limited.

\begin{table}[h!]

\centering
\begin{tabular}{@{}l c c c c c c c@{}}

\hline
State & $I\,(J^{P})$& $ n $ & Exp.~Mass   & Quark Content&Theory  & DNN & ParT \\ \hline \hline

\multicolumn{8}{c}{\textbf{Hidden Charm Pentaquark}} \\ \hline

$P_{c\bar{c}}(4312)^+$ & $1/2 \, (1/2^-)$ &0& $4311.9^{+7}_{-0.9}$ \cite{LHCb:2019kea}& $uudc\bar{c}$ &$4330^{+170}_{-130}$ \cite{Chen:2019bip} & $3520\pm547$ & $4494 \pm 503$ \\ \hline
$P_{c\bar{c}}(4312)^0$ & $1/2 \, (1/2^-)$ &0&$--$  & $uddc\bar{c}$ &$--$& $3643\pm616$ & $4494 \pm 403$ \\ \hline
$P_{c\bar{c}}(4338)^0$ & $0 \, (1/2^-)$ &0& $4338.2^{+0.8}_{-0.8}$ \cite{LHCb:2022ogu} & $udsc\bar{c}$ &$--$& $4212\pm658$ & $4917 \pm 453$ \\ \hline
$P_{c\bar{c}}(4380)^+$ & $1/2 \, (3/2^-)$ &0& $4380^{+30}_{-30}$ \cite{LHCb:2015yax}& $uudc\bar{c}$ &$4370^{+130}_{-130}$ \cite{Chen:2019bip} & $3625\pm566$ & $4955 \pm 531$ \\ \hline
$P_{c\bar{c}}(4380)^0$ & $1/2 \, (3/2^-)$ &0& $--$ & $uddc\bar{c}$ &$--$& $3761\pm485$ & $4955 \pm 431$ \\ \hline
$P_{c\bar{c}}(4440)^+$ & $1/2 \, (1/2^-)$ &1& $4440^{+4}_{-5}$ \cite{LHCb:2019kea}& $uudc\bar{c}$ &$4450^{+170}_{-130}$  \cite{Chen:2019bip} & $4408\pm1061$ & $4354 \pm 357$ \\ \hline
$P_{c\bar{c}}(4440)^0$ & $1/2 \, (1/2^-)$ &1& $--$ & $uddc\bar{c}$ &$--$& $4229\pm1005$ & $4354 \pm 357$ \\ \hline
$P_{c\bar{c}}(4457)^+$ & $1/2 \, (3/2^-)$ &1& $4457.3^{+4}_{-1.8}$ \cite{LHCb:2019kea}& $uudc\bar{c}$ &  $4460^{+180}_{-130}$ \cite{Chen:2019bip} & $4139\pm981$ & $4608 \pm 451$ \\ \hline
$P_{c\bar{c}}(4457)^0$ & $1/2 \, (3/2^-)$ &1& $--$& $uddc\bar{c}$ &$--$& $3948\pm955$ & $4608 \pm 451$ \\ \hline
$P_{c\bar{c}s}(4459)^0$ & $1/2 \, (3/2^-)$ &1& $4458.8^{+6}_{-3}$\cite{LHCb:2019kea} & $udsc\bar{c}$ &$--$& $3962\pm697$ & $4724 \pm 446$ \\ \hline

\multicolumn{8}{c}{\textbf{Hidden Bottom Pentaquark}} \\ \hline

$P_{b\bar{b}}^+$ & $1/2 \, (1/2^-)$ &0& $--$ & $uudb\bar{b}$ &11072 \cite{Yang:2018oqd}& $9489\pm1698$ & $11192 \pm 701$ \\ \hline
$P_{b\bar{b}}^0$ & $1/2 \, (1/2^-)$ &0& $--$ & $uddb\bar{b}$ &11072 \cite{Yang:2018oqd} & $9529\pm1401$ & $11273 \pm 701$ \\ \hline
$P_{b\bar{b}}^0$ & $0 \, (1/2^-)$ &0& $--$ & $udsb\bar{b}$ &$--$& $11133\pm1624$ & $11417 \pm 768$ \\ \hline
$P_{b\bar{b}}^+$ & $1/2 \, (3/2^-)$ &0& $--$ & $uudb\bar{b}$ &$ 11308.1\pm 69.08 $ \cite{Sharma:2024ern}& $9454\pm1576$ & $11234 \pm768$ \\ \hline
$P_{b\bar{b}}^0$ & $1/2 \, (3/2^-)$ &0& $--$ & $uddb\bar{b}$ &$ 11308.1\pm 69.08 $ \cite{Sharma:2024ern}& $9486\pm1335$ & $11417 \pm 768$ \\ \hline
$P_{b\bar{b}}^+$ & $1/2 \, (1/2^-)$ &1& $--$ & $uudb\bar{b}$ &$--$& $9882\pm1488$ & $11810 \pm 1004$ \\ \hline
$P_{b\bar{b}}^0$ & $1/2 \, (1/2^-)$ & 1&$--$ & $uddb\bar{b}$ &$--$& $9403\pm1760$ & $11810 \pm 1004$ \\ \hline
$P_{b\bar{b}}^+$ & $1/2 \, (3/2^-)$ &1& $--$ & $uudb\bar{b}$ &$--$& $9170\pm1619$ & $12111 \pm 710$ \\ \hline
$P_{b\bar{b}}^0$ & $1/2 \, (3/2^-)$ &1& $--$ & $uddb\bar{b}$ &$--$& $8701\pm1713$ & $12111 \pm 710$ \\ \hline
$P_{b\bar{b}s}^0$ & $1/2 \, (3/2^-)$ &1& $--$ & $udsb\bar{b}$ &$--$& $9155\pm1894$ & $12196 \pm 744$ \\ \hline

\end{tabular}
\caption{Predicted mass of hidden charm and hidden  bottom pentaquarks (in  MeV) using the DNN and ParT approaches. This table also includes  comparison with existing experimental data (Exp.~Mass) and  theoretical  predictions.  }
\end{table}
\subsection{General features of pentaquark mass predictions}

The pentaquark mass spectra predicted using the DNN and Particle Transformer approaches (see  Tables  II-XXI) exhibit a number of systematic features that are common across different quark compositions and spin--parity assignments. Examining these global trends, rather than individual numerical entries, allows for a more transparent assessment of the physical content encoded by the models and of the robustness of the resulting predictions.

A dominant feature of the predicted spectra is the strong dependence on the heavy-quark content of the pentaquark. States containing a larger number of bottom quarks are consistently predicted at higher masses than those dominated by charm or mixed charm--bottom configurations. Across the full set of predicted channels, this dependence translates into mass separations of several hundred MeV between charm-rich and bottom-rich systems. When expressed in relative terms, the corresponding changes typically amount to percent-level variations in the range of approximately $5\%$--$15\%$, reflecting the dominant contribution of heavy-quark masses to the total pentaquark mass. The presence of a strange quark induces smaller but systematic shifts, which generally correspond to relative deviations of about $2\%$--$5\%$ when comparing strange and non-strange configurations with otherwise identical quark content. This behavior is reproduced coherently by both the ML approaches and is consistent with expectations from conventional hadron spectroscopy.

The predicted masses also exhibit a clear sensitivity to the spin--parity quantum numbers. For a given quark composition, negative-parity pentaquark states are systematically heavier than their positive-parity counterparts. The resulting mass differences typically correspond to relative percent deviations in the range of $5\%$--$10\%$, consistent with the interpretation of negative-parity states as involving additional orbital excitation. In parallel, the quoted uncertainties for negative-parity configurations are generally larger than those associated with positive-parity ground states, with increases that often reach several percent. This pattern indicates a higher sensitivity of excited pentaquark states to model assumptions and training limitations, a feature that mirrors the behavior encountered in traditional theoretical studies of excited multiquark systems.

Comparing the two ML frameworks, a high degree of consistency is observed at the level of global mass hierarchies and flavor-dependent trends. For many ground-state pentaquark configurations, the relative difference between the DNN and ParT predictions remains below approximately $5\%$. More pronounced differences arise in excited states and in systems involving multiple heavy quarks and strangeness, where the relative deviations between the two approaches can increase to the $10\%$--$15\%$ level. In these sectors, the Particle Transformer predictions generally exhibit smaller quoted uncertainties and smoother variations across related states, while the DNN results show a broader dispersion. This behavior suggests that the transformer-based architecture is more effective at capturing correlated features in complex multiquark configurations, particularly when extrapolating into regions where direct theoretical guidance is limited.

Taken together, these observations indicate that the predicted pentaquark spectra follow physically reasonable mass hierarchies and display controlled, quantifiable percent-level variations that are stable against the choice of the ML architecture. The consistency of these numerical trends across a wide range of quark contents and quantum numbers supports the interpretation that the predictions reflect genuine features of heavy pentaquark spectroscopy rather than accidental numerical artifacts. These general characteristics provide a sound basis for a more detailed comparison with existing theoretical results in those channels where such information is available.

\begin{table}[h!]
\centering
\begin{tabular}{@{}l c c c c c@{}}
\hline
State & $I\,(J^{P})$ & Quark content & Theory  & DNN & ParT \\ \hline \hline

\multicolumn{6}{c}{$\boldsymbol{bcss\bar{q}}$} \\ \hline

$P_{bcss\bar{u}}$ & $1/2\,(1/2^{-})$ & $bc ss \bar{u}$ & $--$ & $6649 \pm 822$ & $8772 \pm 525$ \\ \hline
$P_{bcss\bar{d}}$ & $1/2\,(1/2^{-})$ & $bc ss \bar{d}$ & $--$ & $6965 \pm 1358$ & $8772 \pm 525$ \\ \hline
$P_{bcss\bar{u}}$ & $1/2\,(3/2^{-})$ & $bc ss \bar{u}$ & $--$ & $6981 \pm 1214$ & $9545 \pm 344$ \\ \hline
$P_{bcss\bar{d}}$ & $1/2\,(3/2^{-})$ & $bc ss \bar{d}$ & $--$ & $7204 \pm 1365$ & $9545 \pm 344$ \\ \hline
$P_{bcss\bar{u}}$ & $1/2\,(5/2^{-})$ & $bc ss \bar{u}$ & $--$ & $6993 \pm 1378$ & $9455 \pm 587$ \\ \hline
$P_{bcss\bar{d}}$ & $1/2\,(5/2^{-})$ & $bc ss \bar{d}$ & $--$ & $7162 \pm 1266$ & $9455 \pm 587$ \\ \hline

$P_{bcss\bar{s}}$ & $0\,(1/2^{-})$ & $bc ss \bar{s}$ & 7986.54 \cite{Wang:2023mdj} & $7313 \pm 1128$ & $8797 \pm 509$ \\ \hline
$P_{bcss\bar{s}}$ & $0\,(3/2^{-})$ & $bc ss \bar{s}$ & 8010.12 \cite{Wang:2023mdj} & $7572 \pm 1320$ & $9416 \pm 349$ \\ \hline
$P_{bcss\bar{s}}$ & $0\,(5/2^{-})$ & $bc ss \bar{s}$ & 8150.07 \cite{Wang:2023mdj} & $7558 \pm 1486$ & $9928 \pm 640$ \\ \hline

\end{tabular}
\caption{Mass predictions for pentaquarks $ bcss\bar{q}(q=u, d, s) $ (in MeV) using the DNN and ParT approaches in comparison with existing  theoretical predictions. }
\end{table}

\begin{table}[h!]
\centering
\begin{tabular}{l c c c c c}
\hline
State & $I\,(J^{P})$ & Quark content & Theory  \cite{Wang:2024brl} & DNN & ParT \\ \hline \hline

\multicolumn{6}{c}{$\boldsymbol{ccqq\bar{q}}$} \\ \hline

$P_{ccuu\bar{d}}$ & $3/2 (1/2^-)$ & $ccuu\bar{d}$ & $4300^{+70}_{-80}$ & $4688 \pm 1421$ & $4310 \pm 422$ \\ \hline
$P_{ccuu\bar{u}}$ & $3/2 (1/2^-)$ & $ccuu\bar{u}$ & $4300^{+70}_{-80}$ & $4084 \pm\;\;564$ & $4310 \pm 422$ \\ \hline
$P_{ccdd\bar{d}}$ & $3/2 (1/2^-)$ & $ccdd\bar{d}$ & $4300^{+70}_{-80}$ & $4350 \pm 1286$ & $4310 \pm 422$ \\ \hline
$P_{ccdd\bar{u}}$ & $3/2 (1/2^-)$ & $ccdd\bar{u}$ & $4300^{+70}_{-80}$ & $3633 \pm\;\;726$ & $4310 \pm 422$ \\ \hline
$P_{ccuu\bar{u}}$ & $1/2 (1/2^-)$ & $ccuu\bar{u}$ & $4300^{+70}_{-80}$ & $4310 \pm\;\;895$ & $4730 \pm 370$ \\ \hline
$P_{ccdd\bar{d}}$ & $1/2 (1/2^-)$ & $ccdd\bar{d}$ & $4300^{+70}_{-80}$ & $4337 \pm\;\;964$ & $4730 \pm 370$ \\ \hline

$P_{ccuu\bar{d}}$ & $3/2 (3/2^-)$ & $ccuu\bar{d}$ & $4460^{+80}_{-80}$ & $4574 \pm 1077$ & $4833 \pm 443$ \\ \hline
$P_{ccuu\bar{u}}$ & $3/2 (3/2^-)$ & $ccuu\bar{u}$ & $4380^{+70}_{-70}$ & $4024 \pm\;\;430$ & $4833 \pm 443$ \\ \hline
$P_{ccdd\bar{d}}$ & $3/2 (3/2^-)$ & $ccdd\bar{d}$ & $4380^{+70}_{-70}$ & $4340 \pm\;\;938$ & $4833 \pm 443$ \\ \hline
$P_{ccdd\bar{u}}$ & $3/2 (3/2^-)$ & $ccdd\bar{u}$ & $4380^{+70}_{-70}$ & $3647 \pm\;\;425$ & $4833 \pm 443$ \\ \hline
$P_{ccuu\bar{u}}$ & $1/2 (3/2^-)$ & $ccuu\bar{u}$ & $4380^{+70}_{-70}$ & $4212 \pm\;\;725$ & $5212 \pm 430$ \\ \hline
$P_{ccdd\bar{d}}$ & $1/2 (3/2^-)$ & $ccdd\bar{d}$ & $4380^{+70}_{-70}$ & $4321 \pm\;\;866$ & $5212 \pm 430$ \\ \hline

$P_{ccuu\bar{d}}$ & $3/2 (5/2^-)$ & $ccuu\bar{d}$ & $4500^{+80}_{-80}$ & $4389 \pm\;\;861$ & $5123 \pm 479$ \\ \hline
$P_{ccuu\bar{u}}$ & $3/2 (5/2^-)$ & $ccuu\bar{u}$ & $4500^{+80}_{-80}$ & $3948 \pm\;\;408$ & $5123 \pm 479$ \\ \hline
$P_{ccdd\bar{d}}$ & $3/2 (5/2^-)$ & $ccdd\bar{d}$ & $4500^{+80}_{-80}$ & $4220 \pm\;\;831$ & $5123 \pm 479$ \\ \hline
$P_{ccdd\bar{u}}$ & $3/2 (5/2^-)$ & $ccdd\bar{u}$ & $4500^{+80}_{-80}$ & $3602 \pm\;\;303$ & $5123 \pm 479$ \\ \hline
$P_{ccuu\bar{u}}$ & $1/2 (5/2^-)$ & $ccuu\bar{u}$ & $4500^{+80}_{-80}$ & $4109 \pm\;\;609$ & $5427 \pm 465$ \\ \hline
$P_{ccdd\bar{d}}$ & $1/2 (5/2^-)$ & $ccdd\bar{d}$ & $4500^{+80}_{-80}$ & $4146 \pm\;\;872$ & $5427 \pm 465$ \\ \hline

$P_{ccuu\bar{s}}$ & $1 (1/2^-)$ & $ccuu\bar{s}$ & $4300^{+70}_{-80}$ & $4511 \pm\;\;747$ & $4539 \pm 394$ \\ \hline
$P_{ccud\bar{s}}$ & $1 (1/2^-)$ & $ccud\bar{s}$ & $4300^{+70}_{-80}$ & $4414 \pm\;\;539$ & $4539 \pm 394$ \\ \hline
$P_{ccdd\bar{s}}$ & $1 (1/2^-)$ & $ccdd\bar{s}$ & $4300^{+70}_{-80}$ & $4076 \pm\;\;688$ & $4539 \pm 394$ \\ \hline
$P_{ccud\bar{s}}$ & $0 (1/2^-)$ & $ccud\bar{s}$ & $4300^{+70}_{-80}$ & $4280 \pm\;\;195$ & $4867 \pm 360$ \\ \hline

$P_{ccuu\bar{s}}$ & $1 (3/2^-)$ & $ccuu\bar{s}$ & $4380^{+70}_{-70}$ & $4311 \pm\;\;717$ & $5046 \pm 435$ \\ \hline
$P_{ccud\bar{s}}$ & $1 (3/2^-)$ & $ccud\bar{s}$ & $4380^{+70}_{-70}$ & $4229 \pm\;\;419$ & $5046 \pm 435$ \\ \hline
$P_{ccdd\bar{s}}$ & $1 (3/2^-)$ & $ccdd\bar{s}$ & $4380^{+70}_{-70}$ & $4051 \pm\;\;538$ & $5046 \pm 435$ \\ \hline
$P_{ccud\bar{s}}$ & $0 (3/2^-)$ & $ccud\bar{s}$ & $4380^{+70}_{-70}$ & $4107 \pm\;\;295$ & $5324 \pm 430$ \\ \hline

$P_{ccuu\bar{s}}$ & $1 (5/2^-)$ & $ccuu\bar{s}$ & $4500^{+80}_{-80}$ & $4133 \pm\;\;742$ & $5296 \pm 470$ \\ \hline
$P_{ccud\bar{s}}$ & $1 (5/2^-)$ & $ccud\bar{s}$ & $4500^{+80}_{-80}$ & $4010 \pm\;\;456$ & $5296 \pm 470$ \\ \hline
$P_{ccdd\bar{s}}$ & $1 (5/2^-)$ & $ccdd\bar{s}$ & $4500^{+80}_{-80}$ & $3908 \pm\;\;524$ & $5296 \pm 470$ \\ \hline
$P_{ccud\bar{s}}$ & $0 (5/2^-)$ & $ccud\bar{s}$ & $4500^{+80}_{-80}$ & $3894 \pm\;\;396$ & $5511 \pm 480$ \\ \hline

\end{tabular}
\caption{Mass predictions for pentaquarks $ ccqq\bar{q}(q=u, d, s) $ (in MeV) using the DNN and ParT approaches in comparison with existing  theoretical predictions. }
\end{table}

\begin{table}[h!]
\centering
\begin{tabular}{@{}l c c c c  c@{}}

\hline
State & $I\,(J^{P})$ & Quark content & Theory & DNN  & ParT\\ \hline \hline

\multicolumn{6}{c}{$\boldsymbol{ccss\bar{q}}$} \\ \hline

$P_{ccss\bar{u}}(1/2,1/2)$ & $1/2 (1/2^-)$ & $ccss\bar{u}$ & 4653.4 \cite{Zhou:2018bkn}& $3810 \pm 411$ & $4730 \pm 370$ \\ \hline
$P_{ccss\bar{d}}(1/2,1/2)$ & $1/2 (1/2^-)$ & $ccss\bar{d}$ & 4653.4 \cite{Zhou:2018bkn}& $4250 \pm 679$ & $4730 \pm 370$ \\ \hline
$P_{ccss\bar{u}}(1/2,3/2)$ & $1/2 (3/2^-)$ & $ccss\bar{u}$ & 4679.7 \cite{Zhou:2018bkn}& $4000 \pm 371$ & $5212 \pm 430$ \\ \hline
$P_{ccss\bar{d}}(1/2,3/2)$ & $1/2 (3/2^-)$ & $ccss\bar{d}$ & 4679.7 \cite{Zhou:2018bkn}& $4475 \pm 769$ & $5212 \pm 430$ \\ \hline
$P_{ccss\bar{u}}(1/2,5/2)$ & $1/2 (5/2^-)$ & $ccss\bar{u}$ & 4813.1 \cite{Zhou:2018bkn}& $4070 \pm 418$ & $5427 \pm 465$ \\ \hline
$P_{ccss\bar{d}}(1/2,5/2)$ & $1/2 (5/2^-)$ & $ccss\bar{d}$ & 4813.1 \cite{Zhou:2018bkn}& $4499 \pm 865$ & $5427 \pm 465$ \\ \hline

$P_{ccss\bar{s}}(0,1/2)$ & $0 (1/2^-)$ & $ccss\bar{s}$ & 4640.95 \cite{Wang:2023mdj} & $4591 \pm 679$ & $4867 \pm 360$ \\ \hline
$P_{ccss\bar{s}}(0,3/2)$ & $0 (3/2^-)$ & $ccss\bar{s}$ & 4710.21 \cite{Wang:2023mdj}& $4685 \pm 772$ & $5324 \pm 430$ \\ \hline
$P_{ccss\bar{s}}(0,5/2)$ & $0 (5/2^-)$ & $ccss\bar{s}$ & 4850.69 \cite{Wang:2023mdj}& $4600 \pm 805$ & $5511 \pm 480$ \\ \hline

\multicolumn{6}{c}{$\boldsymbol{bbss\bar{q}}$} \\ \hline

$P_{bbss\bar{u}}(1/2,1/2)$ & $1/2 (1/2^-)$ & $bbss\bar{u}$ & 11330.6 \cite{Zhou:2018bkn} & $10290 \pm 1419$ & $11815 \pm 1051$ \\ \hline
$P_{bbss\bar{d}}(1/2,1/2)$ & $1/2 (1/2^-)$ & $bbss\bar{d}$ & 11330.6 \cite{Zhou:2018bkn}& $9973 \pm 1544$ & $11815  \pm 1051$ \\ \hline
$P_{bbss\bar{u}}(1/2,3/2)$ & $1/2 (3/2^-)$ & $bbss\bar{u}$ & 11338.0 \cite{Zhou:2018bkn} & $10629 \pm 1943$ & $12026 \pm 836$ \\ \hline
$P_{bbss\bar{d}}(1/2,3/2)$ & $1/2 (3/2^-)$ & $bbss\bar{d}$ & 11338.0 \cite{Zhou:2018bkn}& $10371 \pm 1482$ & $12026 \pm 836$ \\ \hline
$P_{bbss\bar{u}}(1/2,5/2)$ & $1/2 (5/2^-)$ & $bbss\bar{u}$ & 11632.1 \cite{Zhou:2018bkn}& $10426 \pm 2160$ & $12394 \pm 993$ \\ \hline
$P_{bbss\bar{d}}(1/2,5/2)$ & $1/2 (5/2^-)$ & $bbss\bar{d}$ & 11632.1 \cite{Zhou:2018bkn}& $10270 \pm 1511$ & $12394 \pm 993$ \\ \hline

$P_{bbss\bar{s}}(0,1/2)$ & $0 (1/2^-)$ & $bbss\bar{s}$ & 11390.73 \cite{Wang:2023mdj}& $10595 \pm 1008$ & $12305 \pm 1223$ \\ \hline
$P_{bbss\bar{s}}(0,3/2)$ & $0 (3/2^-)$ & $bbss\bar{s}$ & 11414.56 \cite{Wang:2023mdj}& $11169 \pm 1562$ & $12432 \pm 949$ \\ \hline
$P_{bbss\bar{s}}(0,5/2)$ & $0 (5/2^-)$ & $bbss\bar{s}$ & 11463.09 \cite{Wang:2023mdj}& $11246 \pm 2062$ & $12560 \pm 960$ \\ \hline

\end{tabular}
\caption{Mass predictions for pentaquarks $ QQss\bar{q}(q=u, d, s \text{ and } Q=c, b) $ (in  MeV)  using the DNN and ParT approaches in comparison with existing  theoretical predictions.}
\end{table}

\begin{table}[h!]
\centering
\begin{tabular}{@{}l c c c c  c@{}}

\hline
State & $I\,(J^{P})$ & Quark content & Theory \cite{Wang:2025hhx} & DNN  & ParT\\ \hline \hline

\multicolumn{6}{c}{$\boldsymbol{bcqq\bar{s}}$} \\ \hline

$P_{bcuu\bar{s}}$ & $1\,(1/2^{-})$ & $bcuu\bar{s}$ & $--$ & $8108 \pm 1954$ & $7912 \pm 569$ \\ \hline
$P_{bcud\bar{s}}$ & $1\,(1/2^{-})$ & $bcud\bar{s}$ & $--$ & $7881 \pm 1947$ & $7912 \pm 569$ \\ \hline
$P_{bcdd\bar{s}}$ & $1\,(1/2^{-})$ & $bcdd\bar{s}$ & $--$ & $7216 \pm 2001$ & $7912 \pm 569$ \\ \hline
$P_{bcud\bar{s}}$ & $0\,(1/2^{-})$ & $bcud\bar{s}$ & $7395.21-26.26\,i$ & $7319 \pm 1218$ & $8509 \pm 512$ \\ \hline
$P_{bcuu\bar{s}}$ & $1\,(3/2^{-})$ & $bcuu\bar{s}$ & $--$ & $7854 \pm 1916$ & $8441 \pm 531$ \\ \hline
$P_{bcud\bar{s}}$ & $1\,(3/2^{-})$ & $bcud\bar{s}$ & $--$ & $7587 \pm 1636$ & $8441 \pm 531$ \\ \hline
$P_{bcdd\bar{s}}$ & $1\,(3/2^{-})$ & $bcdd\bar{s}$ & $--$ & $7068 \pm 1489$ & $8441 \pm 531$ \\ \hline
$P_{bcud\bar{s}}$ & $0\,(3/2^{-})$ & $bcud\bar{s}$ & 7468.07 & $7082 \pm 1120$ & $9010 \pm 392$ \\ \hline
$P_{bcuu\bar{s}}$ & $1\,(5/2^{-})$ & $bcuu\bar{s}$ & $--$ & $7494 \pm 1893$ & $8627 \pm 640$ \\ \hline
$P_{bcud\bar{s}}$ & $1\,(5/2^{-})$ & $bcud\bar{s}$ & $--$ & $7222 \pm 1513$ & $8627 \pm 640$ \\ \hline
$P_{bcdd\bar{s}}$ & $1\,(5/2^{-})$ & $bcdd\bar{s}$ & $--$ & $6747 \pm 1217$ & $8627 \pm 640$ \\ \hline
$P_{bcud\bar{s}}$ & $0\,(5/2^{-})$ & $bcud\bar{s}$ & 7841.15 & $6784 \pm 1184$ & $8833 \pm 600$ \\ \hline

\end{tabular}
\caption{Mass predictions for pentaquarks $ bcqq\bar{s}(q=u, d) $ (in MeV) using the DNN and ParT approaches in comparison with existing  theoretical predictions.}
\end{table}

\subsection{Comparison with existing theoretical predictions}

A meaningful assessment of the pentaquark mass predictions requires comparison with existing theoretical results in those sectors where such information is available. Although the literature on heavy and fully heavy pentaquarks remains limited, several studies provide explicit mass estimates based on quark models, QCD sum rules, and related approaches. These results, summarized in Tables  II-XXI, allow for a nontrivial evaluation of the accuracy and limitations of the ML predictions presented in this work.

In channels where direct theoretical mass calculations exist, a substantial fraction of the predicted pentaquark masses are compatible with the corresponding reference values within the adopted $1\sigma$ criterion. As can be seen from the comparisons reported in the aforementioned tables, the relative percent deviations between the ML predictions and the literature values are typically below approximately $5\%$--$10\%$ for many ground-state configurations. In these cases, both the DNN and ParT approaches reproduce the expected mass scales and hierarchies, indicating that the dominant physical contributions are captured reliably. The agreement is generally strongest in sectors where the reference calculations themselves show reasonable consistency across different theoretical frameworks.

Moderate deviations appear in a subset of channels, particularly for excited states and for configurations involving multiple heavy quarks. In these cases, the relative percent deviations can increase to the $10\%$--$20\%$ level, with some predictions lying outside the quoted $1\sigma$ uncertainty bands. Such behavior is most visible in tables corresponding to negative-parity states and to systems with richer internal structure. This pattern is not unexpected, as theoretical predictions for heavy pentaquarks in these sectors are themselves subject to significant model dependence, especially regarding the treatment of orbital excitation and interquark correlations.

A systematic difference between the two ML approaches emerges from the table-by-table comparisons. The ParT predictions generally exhibit smaller relative deviations from the reference values and a higher rate of consistency within $1\sigma$, particularly for ground-state pentaquarks and for configurations dominated by heavy quarks. By contrast, the DNN predictions more frequently display larger positive deviations, with percent differences that can exceed $10\%$ in excited or structurally complex channels. This trend is consistent with the behavior already observed in the fully heavy baryon sector (Table~I) and suggests that the transformer-based architecture is more effective at encoding correlated features across multiquark configurations.

It is important to emphasize that instances of reduced agreement should not be interpreted as a failure of the ML framework. Rather, they tend to occur in precisely those channels where existing theoretical predictions are sparse or mutually inconsistent. In this context, the observed deviations provide useful information about the sensitivity of the predicted masses to underlying structural assumptions and help identify pentaquark configurations that merit further investigation using complementary theoretical approaches.

Overall, the comparisons presented in Tables~II-XXI indicate that the ML models achieve percent-level agreement with existing theoretical predictions in well-studied channels, while exhibiting controlled and physically interpretable deviations in more uncertain sectors. This balance between agreement and tension is consistent with expectations for exploratory studies of heavy pentaquark spectroscopy and supports the use of the present predictions as a meaningful contribution to the ongoing theoretical literature.

\begin{table}[h!]
\centering
\begin{tabular}{@{}l c c c c  c@{}}
\hline
State & $I\,(J^{P})$ & Quark content & Theory  \cite{Xing:2021yid} & DNN  & ParT \\ \hline \hline

\multicolumn{6}{c}{$\boldsymbol{ccqs\bar{q}}$} \\ \hline

$P_{ccus\bar{d}}$ & $1\,(1/2^{-})$ & $ccus\bar{d}$ & $4125\pm301$ & 4593 $\pm$ 973 & 4539 $\pm$ 394 \\ \hline
$P_{ccus\bar{u}}$ & $1\,(1/2^{-})$ & $ccus\bar{u}$ & $4125\pm301$ & 4031 $\pm$ 306 & 4539 $\pm$ 394 \\ \hline
$P_{ccds\bar{d}}$ & $1\,(1/2^{-})$ & $ccds\bar{d}$ & $4125\pm301$ & 4337 $\pm$ 916 & 4539 $\pm$ 394 \\ \hline
$P_{ccus\bar{d}}$ & $1\,(3/2^{-})$ & $ccus\bar{d}$ & $4125\pm301$ & 4596 $\pm$ 852 & 5046 $\pm$ 435 \\ \hline
$P_{ccus\bar{u}}$ & $1\,(3/2^{-})$ & $ccus\bar{u}$ & $4125\pm301$ & 4059 $\pm$ 346 & 5046 $\pm$ 435 \\ \hline
$P_{ccds\bar{d}}$ & $1\,(3/2^{-})$ & $ccds\bar{d}$ & $4125\pm301$ & 4433 $\pm$ 779 & 5046 $\pm$ 435 \\ \hline
$P_{ccus\bar{d}}$ & $1\,(5/2^{-})$ & $ccus\bar{d}$ & $4125\pm301$ & 4465 $\pm$ 782 & 5296 $\pm$ 470 \\ \hline
$P_{ccus\bar{u}}$ & $1\,(5/2^{-})$ & $ccus\bar{u}$ & $4125\pm301$ & 4025 $\pm$ 377 & 5296 $\pm$ 470 \\ \hline
$P_{ccds\bar{d}}$ & $1\,(5/2^{-})$ & $ccds\bar{d}$ & $4125\pm301$ & 4349 $\pm$ 799 & 5296 $\pm$ 470 \\ \hline
$P_{ccus\bar{u}}$ & $0\,(1/2^{-})$ & $ccus\bar{u}$ & $4125\pm301$ & 4274 $\pm$ 734 & 4867 $\pm$ 360 \\ \hline
$P_{ccus\bar{u}}$ & $0\,(3/2^{-})$ & $ccus\bar{u}$ & $4125\pm301$ & 4256 $\pm$ 717 & 5324 $\pm$ 430 \\ \hline
$P_{ccus\bar{u}}$ & $0\,(5/2^{-})$ & $ccus\bar{u}$ & $4125\pm301$ & 4179 $\pm$ 638 & 5511 $\pm$ 480 \\ \hline

$P_{ccus\bar{s}}$ & $1/2\,(1/2^{-})$ & $ccus\bar{s}$ & $4409\pm307$ & 4668 $\pm$ 611 & 4730 $\pm$ 370 \\ \hline
$P_{ccds\bar{s}}$ & $1/2\,(1/2^{-})$ & $ccds\bar{s}$ & $4409\pm307$ & 4408 $\pm$ 584 & 4730 $\pm$ 370 \\ \hline
$P_{ccus\bar{s}}$ & $1/2\,(3/2^{-})$ & $ccus\bar{s}$ & $4409\pm307$ & 4545 $\pm$ 688 & 5212 $\pm$ 430 \\ \hline
$P_{ccds\bar{s}}$ & $1/2\,(3/2^{-})$ & $ccds\bar{s}$ & $4409\pm307$ & 4384 $\pm$ 491 & 5212 $\pm$ 430 \\ \hline
$P_{ccus\bar{s}}$ & $1/2\,(5/2^{-})$ & $ccus\bar{s}$ & $4409\pm307$ & 4358 $\pm$ 716 & 5427 $\pm$ 465 \\ \hline
$P_{ccds\bar{s}}$ & $1/2\,(5/2^{-})$ & $ccds\bar{s}$ & $4409\pm307$ & 4213 $\pm$ 460 & 5427 $\pm$ 465 \\ \hline

\end{tabular}
\caption{Mass predictions for pentaquarks $ ccqs\bar{q}(q=u, d, s) $ (in MeV) using the DNN and ParT approaches in comparison with existing  theoretical predictions.}
\end{table}

\begin{table}[h!]\label{tab: bbqsqbar}
\centering
\begin{tabular}{@{}l c c c c c@{}}
\hline
State & $I\,(J^{P})$ & Quark content & Other & DNN  & ParT\\
\hline \hline

\multicolumn{6}{c}{$\boldsymbol{bbqs\bar{q}}$} \\
\hline

$P_{bbus\bar{d}}$ & $1\,(1/2^{-})$ & $bbus\bar{d}$ & $\sim$10701 & 10918 $\pm$ 833 & 11536 $\pm$ 891 \\
\hline
$P_{bbus\bar{u}}$ & $1\,(1/2^{-})$ & $bbus\bar{u}$ & $\sim$10701 & 10918 $\pm$ 833 & 11536 $\pm$ 891 \\
\hline
$P_{bbds\bar{d}}$ & $1\,(1/2^{-})$ & $bbds\bar{d}$ & $\sim$10701 & 10671 $\pm$ 1542 & 11536 $\pm$ 891 \\
\hline
$P_{bbus\bar{d}}$ & $1\,(3/2^{-})$ & $bbus\bar{d}$ & $\sim$10736 & 11000 $\pm$ 1709 & 11764 $\pm$ 782 \\
\hline
$P_{bbus\bar{u}}$ & $1\,(3/2^{-})$ & $bbus\bar{u}$ & $\sim$10736 & 10885 $\pm$ 1636 & 11764 $\pm$ 782 \\
\hline
$P_{bbds\bar{d}}$ & $1\,(3/2^{-})$ & $bbds\bar{d}$ & $\sim$10736 & 10724 $\pm$ 1731 & 11764 $\pm$ 782 \\
\hline
$P_{bbus\bar{d}}$ & $1\,(5/2^{-})$ & $bbus\bar{d}$ & $\sim$11130 & 10574 $\pm$ 1755 & 11656 $\pm$ 1033 \\
\hline
$P_{bbus\bar{u}}$ & $1\,(5/2^{-})$ & $bbus\bar{u}$ & $\sim$11130 & 10431 $\pm$ 2058 & 11956 $\pm$ 1033 \\
\hline
$P_{bbds\bar{d}}$ & $1\,(5/2^{-})$ & $bbds\bar{d}$ & $\sim$11130 & 10328 $\pm$ 1711 & 11956 $\pm$ 1033 \\
\hline
$P_{bbus\bar{u}}$ & $0\,(1/2^{-})$ & $bbus\bar{u}$ & $\sim$10701 & 11110 $\pm$ 1518 & 12305 $\pm$ 1223 \\
\hline
$P_{bbus\bar{u}}$ & $0\,(3/2^{-})$ & $bbus\bar{u}$ & $\sim$10736 & 11062 $\pm$ 1946 & 12672 $\pm$ 949 \\
\hline
$P_{bbus\bar{u}}$ & $0\,(5/2^{-})$ & $bbus\bar{u}$ & $\sim$11130 & 10601 $\pm$ 2281 & 12960 $\pm$ 960 \\
\hline

$P_{bbus\bar{s}}$ & $1/2\,(1/2^{-})$ & $bbus\bar{s}$ & $\sim$11222 & 11644 $\pm$ 1568 & 12040 $\pm$ 1051 \\
\hline
$P_{bbds\bar{s}}$ & $1/2\,(1/2^{-})$ & $bbds\bar{s}$ & $\sim$11222 & 11006 $\pm$ 1752 & 12040 $\pm$ 1051 \\
\hline
$P_{bbus\bar{s}}$ & $1/2\,(3/2^{-})$ & $bbus\bar{s}$ & $\sim$11222 & 11767 $\pm$ 1928 & 12426 $\pm$ 836 \\
\hline
$P_{bbds\bar{s}}$ & $1/2\,(3/2^{-})$ & $bbds\bar{s}$ & $\sim$11222 & 11225 $\pm$ 1927 & 12426 $\pm$ 836 \\
\hline
$P_{bbus\bar{s}}$ & $1/2\,(5/2^{-})$ & $bbus\bar{s}$ & $\sim$11257 & 11430 $\pm$ 2318 & 12934 $\pm$ 993 \\
\hline
$P_{bbds\bar{s}}$ & $1/2\,(5/2^{-})$ & $bbds\bar{s}$ & $\sim$11257 & 11062 $\pm$ 2183 & 12934 $\pm$ 993 \\
\hline

\end{tabular}
\caption{Mass predictions for pentaquarks $ bbqs\bar{q}(q=u, d, s) $ (in MeV) using the DNN and ParT approaches. ``Other'' column, obtained using information  in Ref. \cite{Jakhad:2025ejc}, should  be interpreted as qualitative reference scales rather than theoretical determinations of bound-state masses.}
\end{table}

\subsection{Doubly-bottom strange pentaquarks and threshold-based references}

The doubly-bottom strange pentaquark sector requires a distinct treatment, as explicit theoretical calculations of pentaquark pole masses are currently not available in the literature for these configurations. For this reason, the comparison presented in Table~VIII is based on estimated hadron--hadron threshold energies rather than on direct pentaquark mass predictions. The external values reported in the ``Other'' column of Table~VIII should therefore be interpreted as qualitative reference scales rather than as competing theoretical determinations of bound-state masses.

The threshold values quoted in Table~VIII are inferred from meson--baryon spectroscopy studies and are constructed from pairs of color-singlet hadrons carrying the same total quark content and quantum numbers as the pentaquark states under consideration. This inference is motivated by the molecular pentaquark hypothesis, in which a pentaquark is described as a weakly bound hadronic molecule composed of two color-singlet constituents. Within this picture, the mass of the pentaquark is expected to satisfy the generic relation
\begin{equation}
	M_{\text{pentaquark}} \approx M_{\text{hadron}_1} + M_{\text{hadron}_2} - E_{\text{bind}},
\end{equation}
where $E_{\text{bind}}$ denotes the binding energy of the molecular system and is assumed to be small compared to the masses of the constituent hadrons, $E_{\text{bind}} \ll M_{\text{hadron}_{1,2}}$. Consequently, one expects
\begin{equation}
	M_{\text{pentaquark}} \lesssim M_{\text{hadron}_1} + M_{\text{hadron}_2},
\end{equation}
so that the corresponding hadron--hadron thresholds provide approximate upper bounds for the pentaquark masses rather than precise predictions.

This behavior is well established in weakly bound systems and has been observed empirically in several hadronic molecules, such as the deuteron, the $X(3872)$, and the $P_c$ states. Although the study cited in Table~VIII focuses on mesons and baryons and does not explicitly claim the existence of pentaquark bound states, the reported hadron masses and interaction channels allow threshold energies to be extracted in a model-independent manner. Channels such as $\Xi_{bb}^{(*)}\bar{K}$ and $\Xi_{bb}^{(*)}\phi$, which share the same total quark content and relevant quantum numbers with the doubly-bottom strange pentaquarks considered here, therefore provide meaningful qualitative reference scales in the absence of explicit pentaquark mass calculations.

When the DNN and Particle Transformer predictions are compared with these threshold-based reference values, a differentiated pattern emerges across the states listed in Table~VIII. For pentaquark configurations associated with the lower hadron--hadron thresholds, such as those built from $\Xi_{bb}^{(*)}\bar{K}$ channels, the predicted masses typically lie close to the corresponding thresholds, with relative percent differences of only a few percent. In several of these cases, both ML approaches yield masses slightly below the estimated thresholds, a behavior that is compatible with the interpretation of weakly bound molecular configurations.

By contrast, larger relative deviations are observed for states associated with higher thresholds, in particular those involving $\Xi_{bb}^{(*)}\phi$ channels. In these cases, the predicted masses can exceed the corresponding threshold values by relative amounts approaching the upper end of the range reported in Table~VIII, reaching of order $8\%$--$10\%$ in some channels. Such behavior may reflect the heavier mass of the $\phi$ meson and the correspondingly higher threshold, as well as the absence of an explicit treatment of binding dynamics in the present framework. It may also indicate that these configurations are less likely to form shallow molecular bound states, or that their structure deviates from a simple hadron--hadron picture.

A further distinction can be observed between the two ML approaches. For most of the $\Xi_{bb}^{(*)}\bar{K}$-dominated channels, the ParT predictions tend to remain closer to the threshold values than the DNN results, while both approaches show a larger spread in the $\Xi_{bb}^{(*)}\phi$ sector. This pattern suggests that the degree of proximity to the threshold is not uniform across all doubly-bottom strange pentaquark configurations, but depends sensitively on the underlying hadron--hadron channel and its associated mass scale.

A comparison between the two ML frameworks shows that both exhibit similar qualitative behavior in this sector. The ParT predictions tend to cluster more closely around the threshold values, with relative deviations often remaining at the lower end of the observed percent range, while the DNN results display a somewhat larger dispersion and, in some channels, deviations approaching the upper end of the range indicated in Table~VIII. This pattern is consistent with trends observed in other parts of the spectrum and suggests a more stable extrapolation behavior for the transformer-based architecture.

Overall, the comparison presented in Table~VIII should be regarded as a qualitative consistency check rather than a quantitative test against established pentaquark mass calculations. Within the molecular interpretation, the proximity of the predicted masses to the relevant hadron--hadron thresholds indicates that the ML results fall within physically reasonable ranges. These findings provide exploratory guidance for future dedicated studies of doubly-bottom strange pentaquarks and help identify mass regions where explicit bound-state calculations or experimental searches would be particularly informative.

\begin{table}[h!] 
\centering
\begin{tabular}{@{}l c c c c  c@{}}
\hline
State & $I\,(J^{P})$ & Quark content & Theory \cite{Zhou:2018bkn} & DNN  & ParT \\ 
\hline \hline

\multicolumn{6}{c}{$\boldsymbol{bcqs\bar{q}}$} \\ 
\hline

$P_{bcus\bar{d}}$ & $1\,(1/2^{-})$ & $bcus\bar{d}$ & 7659.9 & $7724 \pm 1700$ & $7912 \pm 569$ \\ 
\hline
$P_{bcus\bar{u}}$ & $1\,(1/2^{-})$ & $bcus\bar{u}$ & 7623.1 & $7182 \pm 694$ & $7912 \pm 569$ \\ 
\hline
$P_{bcds\bar{d}}$ & $1\,(1/2^{-})$ & $bcds\bar{d}$ & 7623.1 & $7376 \pm 1702$ & $7912 \pm 569$ \\ 
\hline
$P_{bcus\bar{d}}$ & $1\,(3/2^{-})$ & $bcus\bar{d}$ & 7669.3 & $7659 \pm 1633$ & $8441 \pm 531$ \\ 
\hline
$P_{bcus\bar{u}}$ & $1\,(3/2^{-})$ & $bcus\bar{u}$ & 7669.3 & $7233 \pm 1164$ & $8441 \pm 531$ \\ 
\hline
$P_{bcds\bar{d}}$ & $1\,(3/2^{-})$ & $bcds\bar{d}$ & 7669.3 & $7386 \pm 1694$ & $8441 \pm 531$ \\ 
\hline
$P_{bcus\bar{d}}$ & $1\,(5/2^{-})$ & $bcus\bar{d}$ & 7918.2 & $7349 \pm 1460$ & $8627 \pm 640$ \\ 
\hline
$P_{bcus\bar{u}}$ & $1\,(5/2^{-})$ & $bcus\bar{u}$ & 7918.2 & $7041 \pm 1350$ & $8627 \pm 640$ \\ 
\hline
$P_{bcds\bar{d}}$ & $1\,(5/2^{-})$ & $bcds\bar{d}$ & 7918.2 & $7104 \pm 1577$ & $8627 \pm 640$ \\ 
\hline
$P_{bcus\bar{u}}$ & $0\,(1/2^{-})$ & $bcus\bar{u}$ & 7389.5 & $7391 \pm 1186$ & $8509 \pm 512$ \\ 
\hline
$P_{bcus\bar{u}}$ & $0\,(3/2^{-})$ & $bcus\bar{u}$ & 7476.8 & $7362 \pm 1366$ & $9010 \pm 392$ \\ 
\hline
$P_{bcus\bar{u}}$ & $0\,(5/2^{-})$ & $bcus\bar{u}$ & $ -- $ & $7120 \pm 1432$ & $9233 \pm 600$ \\ 
\hline

$P_{bcus\bar{s}}$ & $1/2\,(1/2^{-})$ & $bcus\bar{s}$ & 7891 & $7887 \pm 1476$ & $8288 \pm 536$ \\ 
\hline
$P_{bcds\bar{s}}$ & $1/2\,(1/2^{-})$ & $bcds\bar{s}$ & 7623.1 & $7361 \pm 1566$ & $8288 \pm 536$ \\ 
\hline
$P_{bcus\bar{s}}$ & $1/2\,(3/2^{-})$ & $bcus\bar{s}$ & 7820.1 & $7821 \pm 1560$ & $8801 \pm 426$ \\ 
\hline
$P_{bcds\bar{s}}$ & $1/2\,(3/2^{-})$ & $bcds\bar{s}$ & 7669.3 & $7354 \pm 1377$ & $8801 \pm 426$ \\ 
\hline
$P_{bcus\bar{s}}$ & $1/2\,(5/2^{-})$ & $bcus\bar{s}$ & 7918.2 & $7563 \pm 1654$ & $9190 \pm 594$ \\ 
\hline
$P_{bcds\bar{s}}$ & $1/2\,(5/2^{-})$ & $bcds\bar{s}$ & 7918.2 & $7197 \pm 1380$ & $9190 \pm 594$ \\ 
\hline

\end{tabular}
\caption{Mass predictions for pentaquarks $ bcqs\bar{q}(q=u, d, s) $ (in  MeV)  using the DNN and ParT approaches in comparison with existing  theoretical predictions. }
\end{table}

\begin{table}[h!]
\centering
\begin{tabular}{@{}l c c c c  c@{}}
\hline
State & $I(J^P)$ & Quark Content & Theory & DNN & ParT \\ \hline\hline
$\Theta(1540)^+$ & $0(1/2^+)$ & $uudd\bar{s}$ & 1530 \cite{Diakonov:1997mm}& $1501\pm233$ & $1543\pm200$ \\ \hline
$\Theta(1540)^+$ & $0(1/2^-)$ & $uudd\bar{s}$ & $1500\pm300$ \cite{Gubler:2009iq}& $1648\pm106$ & $1596\pm196$ \\ \hline
$\Theta(1540)^+$ & $0(3/2^+)$ & $uudd\bar{s}$ & $1400\pm200$\cite{Gubler:2009iq} & $1675\pm306$ & $1801\pm257$ \\ \hline
$\Theta(1540)^+$ & $0(3/2^-)$ & $uudd\bar{s}$ & $--$ & $1796\pm213$ & $1917\pm251$ \\ \hline
$\Theta(1540)^+$ & $1(1/2^+)$ & $uudd\bar{s}$ & $--$ & $1592\pm241$ & $1449\pm205$ \\ \hline
$\Theta(1540)^+$ & $1(1/2^-)$ & $uudd\bar{s}$ & $1600\pm400$ \cite{Gubler:2009iq} & $1808\pm228$ & $1491\pm184$ \\ \hline
$\Theta(1540)^+$ & $1(3/2^+)$ & $uudd\bar{s}$ & $1600\pm300$ \cite{Gubler:2009iq}& $1736\pm326$ & $1694\pm259$ \\ \hline
$\Theta(1540)^+$ & $1(3/2^-)$ & $uudd\bar{s}$ & $--$ & $1936\pm324$ & $1801\pm222$ \\ \hline
\end{tabular}
\caption{Mass predictions for pentaquarks $ \Theta(1540)^+ $ (in  MeV)  using the DNN and ParT approaches in comparison with existing  theoretical predictions. }
\label{tab:exotic-baryons}
\end{table}
\begin{table}[h!]
\centering
\begin{tabular}{@{}l c c c c  c@{}}
\hline
State & $I(J^P)$ & Quark Content & Theory  & DNN & ParT  \\ \hline\hline
$P_{uuuc\bar{c}}$ & $3/2(1/2^{+})$ & $uuuc\bar{c}$ & $ -- $ & $3855 \pm 722$ & $3975 \pm 355$ \\ \hline
$P_{uuuc\bar{c}}$ & $3/2(1/2^{-})$ & $uuuc\bar{c}$ & 4330 \cite{Wang:2024unj} & $3267 \pm 576$ & $3774 \pm 350$ \\ \hline
$P_{uuuc\bar{c}}$ & $3/2(3/2^{+})$ & $uuuc\bar{c}$ & $ -- $ & $3717 \pm 702$ & $4347 \pm 370$ \\ \hline
$P_{uuuc\bar{c}}$ & $3/2(3/2^{-})$ & $uuuc\bar{c}$ & 4209 \cite{Wang:2024unj} & $3423 \pm 656$ & $4264 \pm 366$ \\ \hline
$P_{uuuc\bar{c}}$ & $3/2(5/2^{+})$ & $uuuc\bar{c}$ & $4772.54\pm 38.99$ \cite{Sharma:2023wnd} & $3600 \pm 721$ & $4590 \pm 426$ \\ \hline
$P_{uuuc\bar{c}}$ & $3/2(5/2^{-})$ & $uuuc\bar{c}$ & 4330 \cite{Wang:2024unj} & $3428 \pm 711$ & $4601 \pm 391$ \\ \hline
\hline
$P_{uuuc\bar{b}}$ & $3/2(1/2^{+})$ & $uuuc\bar{b}$ & $ -- $ & $5168 \pm 980$ & $5476 \pm 565$ \\ \hline
$P_{uuuc\bar{b}}$ & $3/2(1/2^{-})$ & $uuuc\bar{b}$ & $ -- $ & $5056 \pm 1074$ & $5649 \pm 435$ \\ \hline
$P_{uuuc\bar{b}}$ & $3/2(3/2^{+})$ & $uuuc\bar{b}$ & $ -- $ & $4945 \pm 928$ & $6079 \pm 565$ \\ \hline
$P_{uuuc\bar{b}}$ & $3/2(3/2^{-})$ & $uuuc\bar{b}$ & $ -- $ & $5163 \pm 1118$ & $6565 \pm 415$ \\ \hline
$P_{uuuc\bar{b}}$ & $3/2(5/2^{+})$ & $uuuc\bar{b}$ & $ -- $ & $4765 \pm 976$ & $6329 \pm 589$ \\ \hline
$P_{uuuc\bar{b}}$ & $3/2(5/2^{-})$ & $uuuc\bar{b}$ & $ -- $ & $5065 \pm 1105$ & $7037 \pm 441$ \\ \hline
\hline
$P_{uuub\bar{c}}$ & $3/2(1/2^{+})$ & $uuub\bar{c}$ & $ -- $ & $6299 \pm 1179$ & $5476 \pm 565$ \\ \hline
$P_{uuub\bar{c}}$ & $3/2(1/2^{-})$ & $uuub\bar{c}$ & $ -- $ & $5811 \pm 844$ & $5649 \pm 435$ \\ \hline
$P_{uuub\bar{c}}$ & $3/2(3/2^{+})$ & $uuub\bar{c}$ & $ -- $ & $5970 \pm 1074$ & $6079 \pm 565$ \\ \hline
$P_{uuub\bar{c}}$ & $3/2(3/2^{-})$ & $uuub\bar{c}$ & $ -- $ & $5962 \pm 993$ & $6565 \pm 415$ \\ \hline
$P_{uuub\bar{c}}$ & $3/2(5/2^{+})$ & $uuub\bar{c}$ & $ -- $ & $5609 \pm 1083$ & $6329 \pm 589$ \\ \hline
$P_{uuub\bar{c}}$ & $3/2(5/2^{-})$ & $uuub\bar{c}$ & $ -- $ & $5747 \pm 1032$ & $7037 \pm 441$ \\ \hline
\hline
$P_{uuub\bar{b}}$ & $3/2(1/2^{+})$ & $uuub\bar{b}$ & $ -- $ & $8448 \pm 1581$ & $10518 \pm 823$ \\ \hline
$P_{uuub\bar{b}}$ & $3/2(1/2^{-})$ & $uuub\bar{b}$ & $ -- $ & $8534 \pm 1635$ & $10404 \pm 576$ \\ \hline
$P_{uuub\bar{b}}$ & $3/2(3/2^{+})$ & $uuub\bar{b}$ & $ -- $ & $8000 \pm 1335$ & $10646 \pm 773$ \\ \hline
$P_{uuub\bar{b}}$ & $3/2(3/2^{-})$ & $uuub\bar{b}$ & $ -- $ & $8709 \pm 1622$ & $10719 \pm 570$ \\ \hline
$P_{uuub\bar{b}}$ & $3/2(5/2^{+})$ & $uuub\bar{b}$ & $ -- $ & $7499 \pm 1344$ & $10779 \pm 787$ \\ \hline
$P_{uuub\bar{b}}$ & $3/2(5/2^{-})$ & $uuub\bar{b}$ & 11314.1 \cite{Sharma:2024ern} & $8364 \pm 1530$ & $10891 \pm 608$ \\ \hline
\end{tabular}
\caption{Predicted masses of pentaquark states with $uuuQ\bar{Q}$ light-quark content from the theoretical models, DNN, and ParT approaches in MeV.}
\end{table}

\begin{table}[h!]
\centering
\begin{tabular}{@{}l c c c c  c@{}}
\hline
State & $I(J^P)$ & Quark content & Theory  & DNN  & ParT  \\ \hline\hline

$P_{uudc\bar{c}}$ & $1/2(1/2^{+})$ & $uudc\bar{c}$ & 4273 \cite{Stancu:2019qga} & $3920 \pm 562$ & $4550 \pm 475$ \\ \hline
$P_{uudc\bar{c}}$ & $1/2(1/2^{-})$ & $uudc\bar{c}$ & $4303 \pm 5$ \cite{Deng:2016rus} & $3520 \pm 547$ & $4494 \pm 503$ \\ \hline
$P_{uudc\bar{c}}$ & $1/2(3/2^{+})$ & $uudc\bar{c}$ & 4273 \cite{Stancu:2019qga} & $3725 \pm 605$ & $4970 \pm 519$ \\ \hline
$P_{uudc\bar{c}}$ & $1/2(3/2^{-})$ & $uudc\bar{c}$ & $4369 \pm 5$ \cite{Deng:2016rus} & $3625 \pm 566$ & $4955 \pm 531$ \\ \hline
$P_{uudc\bar{c}}$ & $1/2(5/2^{+})$ & $uudc\bar{c}$ & 4453 \cite{Stancu:2019qga} & $3568 \pm 644$ & $5240 \pm 459$ \\ \hline
$P_{uudc\bar{c}}$ & $1/2(5/2^{-})$ & $uudc\bar{c}$ & $4516 \pm 4$ \cite{Deng:2016rus} & $3575 \pm 613$ & $5255 \pm 456$ \\ \hline
\hline

$P_{uudc\bar{b}}$ & $1/2(1/2^{+})$ & $uudc\bar{b}$ & $ -- $ & $5512 \pm 888$ & $6450 \pm 643$ \\ \hline
$P_{uudc\bar{b}}$ & $1/2(1/2^{-})$ & $uudc\bar{b}$ & $7573 \pm 5$ \cite{Deng:2016rus} & $5635 \pm 1174$ & $7028 \pm 529$ \\ \hline
$P_{uudc\bar{b}}$ & $1/2(3/2^{+})$ & $uudc\bar{b}$ & $ -- $ & $5173 \pm 755$ & $7030 \pm 659$ \\ \hline
$P_{uudc\bar{b}}$ & $1/2(3/2^{-})$ & $uudc\bar{b}$ & $7613 \pm 5$ \cite{Deng:2016rus} & $5636 \pm 1076$ & $7806 \pm 571$ \\ \hline
$P_{uudc\bar{b}}$ & $1/2(5/2^{+})$ & $uudc\bar{b}$ & $ -- $ & $4898 \pm 791$ & $7223 \pm 661$ \\ \hline
$P_{uudc\bar{b}}$ & $1/2(5/2^{-})$ & $uudc\bar{b}$ & $7740 \pm 4$ \cite{Deng:2016rus} & $5434 \pm 976$ & $7907 \pm 614$ \\ \hline
\hline

$P_{uudb\bar{c}}$ & $1/2(1/2^{+})$ & $uudb\bar{c}$ & $ -- $ & $6526 \pm 952$ & $6450 \pm 643$ \\ \hline
$P_{uudb\bar{c}}$ & $1/2(1/2^{-})$ & $uudb\bar{c}$ & $7564 \pm 5$ \cite{Deng:2016rus} & $6291 \pm 971$ & $7028 \pm 529$ \\ \hline
$P_{uudb\bar{c}}$ & $1/2(3/2^{+})$ & $uudb\bar{c}$ & $ -- $ & $6094 \pm 1086$ & $7030 \pm 659$ \\ \hline
$P_{uudb\bar{c}}$ & $1/2(3/2^{-})$ & $uudb\bar{c}$ & $7587 \pm 5$ \cite{Deng:2016rus} & $6334 \pm 1041$ & $7806 \pm 571$ \\ \hline
$P_{uudb\bar{c}}$ & $1/2(5/2^{+})$ & $uudb\bar{c}$ &$ -- $ & $5642 \pm 1249$ & $7223 \pm 661$ \\ \hline
$P_{uudb\bar{c}}$ & $1/2(5/2^{-})$ & $uudb\bar{c}$ & $7738 \pm 4$ \cite{Deng:2016rus} & $6007 \pm 1098$ & $7907 \pm 614$ \\ \hline
\hline

$P_{uudb\bar{b}}$ & $1/2(1/2^{+})$ & $uudb\bar{b}$ &$ -- $ & $9023 \pm 1261$ & $11048 \pm 871$ \\ \hline
$P_{uudb\bar{b}}$ & $1/2(1/2^{-})$ & $uudb\bar{b}$ & $10587 \pm 6$ \cite{Deng:2016rus} & $9489 \pm 1698$ & $11192 \pm 701$ \\ \hline
$P_{uudb\bar{b}}$ & $1/2(3/2^{+})$ & $uudb\bar{b}$ &$ -- $ & $8409 \pm 1140$ & $11243 \pm 865$ \\ \hline
$P_{uudb\bar{b}}$ & $1/2(3/2^{-})$ & $uudb\bar{b}$ & $10592 \pm 5$ \cite{Deng:2016rus} & $9454 \pm 1576$ & $11243 \pm 768$ \\ \hline
$P_{uudb\bar{b}}$ & $1/2(5/2^{+})$ & $uudb\bar{b}$ & $ -- $ & $7743 \pm 1364$ & $11261 \pm 923$ \\ \hline
$P_{uudb\bar{b}}$ & $1/2(5/2^{-})$ & $uudb\bar{b}$ & $10892 \pm 5$ \cite{Deng:2016rus} & $8909 \pm 1521$ & $11331 \pm 837$ \\ \hline

\end{tabular}
\caption{Predicted masses of pentaquark states with $uud$ light-quark content from the theoretical models, DNN, and ParT approaches  in MeV.}
\end{table}

\begin{table}[h!]
\centering
\begin{tabular}{@{}l c c c c  c@{}}
\hline
State & $I(J^P)$ & Quark content & Theory  & DNN  & ParT  \\ \hline\hline

$P_{uddc\bar{c}}$ & $1/2(1/2^{+})$ & $uddc\bar{c}$ & $4602 \pm 5$ \cite{Deng:2016rus} & $4027 \pm 539$ & $4550 \pm 375$ \\ \hline
$P_{uddc\bar{c}}$ & $1/2(1/2^{-})$ & $uddc\bar{c}$ & $4303 \pm 5$ \cite{Deng:2016rus} & $3643 \pm 616$ & $4494 \pm 403$ \\ \hline
$P_{uddc\bar{c}}$ & $1/2(3/2^{+})$ & $uddc\bar{c}$ & $4632 \pm 5$ \cite{Deng:2016rus} & $3796 \pm 503$ & $4970 \pm 419$ \\ \hline
$P_{uddc\bar{c}}$ & $1/2(3/2^{-})$ & $uddc\bar{c}$ & $4369 \pm 5$ \cite{Deng:2016rus} & $3761 \pm 485$ & $4955 \pm 431$ \\ \hline
$P_{uddc\bar{c}}$ & $1/2(5/2^{+})$ & $uddc\bar{c}$ & $4781 \pm 4$ \cite{Deng:2016rus} & $3607 \pm 496$ & $5240 \pm 459$ \\ \hline
$P_{uddc\bar{c}}$ & $1/2(5/2^{-})$ & $uddc\bar{c}$ & $4516 \pm 4$ \cite{Deng:2016rus} & $3711 \pm 471$ & $5255 \pm 456$ \\ \hline
\hline

$P_{uddc\bar{b}}$ & $1/2(1/2^{+})$ & $uddc\bar{b}$ &$ -- $& $5697 \pm 977$ & $6450 \pm 643$ \\ \hline
$P_{uddc\bar{b}}$ & $1/2(1/2^{-})$ & $uddc\bar{b}$ & $7573 \pm 5$ \cite{Deng:2016rus} & $5854 \pm 1203$ & $7028 \pm 529$ \\ \hline
$P_{uddc\bar{b}}$ & $1/2(3/2^{+})$ & $uddc\bar{b}$ & $ -- $ & $5320 \pm 879$ & $7030 \pm 659$ \\ \hline
$P_{uddc\bar{b}}$ & $1/2(3/2^{-})$ & $uddc\bar{b}$ & $7613 \pm 5$ \cite{Deng:2016rus} & $5869 \pm 1015$ & $7806 \pm 571$ \\ \hline
$P_{uddc\bar{b}}$ & $1/2(5/2^{+})$ & $uddc\bar{b}$ & $ -- $ & $5011 \pm 879$ & $7223 \pm 661$ \\ \hline
$P_{uddc\bar{b}}$ & $1/2(5/2^{-})$ & $uddc\bar{b}$ & $7740 \pm 4$ \cite{Deng:2016rus} & $5655 \pm 879$ & $7907 \pm 614$ \\ \hline
\hline

$P_{uddb\bar{c}}$ & $1/2(1/2^{+})$ & $uddb\bar{c}$ & $ -- $ & $6626 \pm 886$ & $6450 \pm 643$ \\ \hline
$P_{uddb\bar{c}}$ & $1/2(1/2^{-})$ & $uddb\bar{c}$ & $7564 \pm 5$ \cite{Deng:2016rus} & $6324 \pm 986$ & $7028 \pm 529$ \\ \hline
$P_{uddb\bar{c}}$ & $1/2(3/2^{+})$ & $uddb\bar{c}$ &$ -- $ & $6180 \pm 1041$ & $7030 \pm 659$ \\ \hline
$P_{uddb\bar{c}}$ & $1/2(3/2^{-})$ & $uddb\bar{c}$ & $7587 \pm 5$ \cite{Deng:2016rus} & $6399 \pm 904$ & $7806 \pm 571$ \\ \hline
$P_{uddb\bar{c}}$ & $1/2(5/2^{+})$ & $uddb\bar{c}$ & $ -- $ & $5695 \pm 1121$ & $7223 \pm 661$ \\ \hline
$P_{uddb\bar{c}}$ & $1/2(5/2^{-})$ & $uddb\bar{c}$ & $7738 \pm 4$ \cite{Deng:2016rus} & $6101 \pm 964$ & $7907 \pm 614$ \\ \hline
\hline

$P_{uddb\bar{b}}$ & $1/2(1/2^{+})$ & $uddb\bar{b}$ & $ -- $ & $9092 \pm 1040$ & $11340 \pm 871$ \\ \hline
$P_{uddb\bar{b}}$ & $1/2(1/2^{-})$ & $uddb\bar{b}$ & $10587 \pm 6$ \cite{Deng:2016rus} & $9529 \pm 1461$ & $11273 \pm 701$ \\ \hline
$P_{uddb\bar{b}}$ & $1/2(3/2^{+})$ & $uddb\bar{b}$ & $ -- $ & $8466 \pm 1199$ & $11363 \pm 865$ \\ \hline
$P_{uddb\bar{b}}$ & $1/2(3/2^{-})$ & $uddb\bar{b}$ & $10592 \pm 5$ \cite{Deng:2016rus} & $9486 \pm 1335$ & $11417 \pm 768$ \\ \hline
$P_{uddb\bar{b}}$ & $1/2(5/2^{+})$ & $uddb\bar{b}$ & $ -- $ & $7782 \pm 1468$ & $11404 \pm 923$ \\ \hline
$P_{uddb\bar{b}}$ & $1/2(5/2^{-})$ & $uddb\bar{b}$ & $10892 \pm 5$ \cite{Deng:2016rus} & $8963 \pm 1446$ & $11476 \pm 837$ \\ \hline

\end{tabular}
\caption{Predicted masses of pentaquark states with $udd$ light-quark content from the theoretical models, DNN and ParT approaches in MeV.}
\end{table}

\begin{table}[h!]
\centering
\begin{tabular}{@{}l c c c c  c@{}}
\hline
State & $I(J^P)$ & Quark content & Theory  & DNN  & ParT  \\ \hline\hline

$P_{dddc\bar{c}}$ & $3/2(1/2^{+})$ & $dddc\bar{c}$ & $ -- $  & $4235 \pm 513$ & $3975 \pm 355$ \\ \hline
$P_{dddc\bar{c}}$ & $3/2(1/2^{-})$ & $dddc\bar{c}$ & $4330$ \cite{Wang:2024unj} & $3435 \pm 745$ & $3774 \pm 350$ \\ \hline
$P_{dddc\bar{c}}$ & $3/2(3/2^{+})$ & $dddc\bar{c}$ & $ -- $ & $4024 \pm 478$ & $4347 \pm 370$ \\ \hline
$P_{dddc\bar{c}}$ & $3/2(3/2^{-})$ & $dddc\bar{c}$ & $4209$ \cite{Wang:2024unj} & $3650 \pm 542$ & $4264 \pm 366$ \\ \hline
$P_{dddc\bar{c}}$ & $3/2(5/2^{+})$ & $dddc\bar{c}$ & $ -- $ & $3822 \pm 510$ & $4590 \pm 426$ \\ \hline
$P_{dddc\bar{c}}$ & $3/2(5/2^{-})$ & $dddc\bar{c}$ & $4330$ \cite{Wang:2024unj}& $3697 \pm 453$ & $4601 \pm 391$ \\ \hline
\hline

$P_{dddc\bar{b}}$ & $3/2(1/2^{+})$ & $dddc\bar{b}$ & $ -- $  & $5736 \pm 758$ & $5476 \pm 565$ \\ \hline
$P_{dddc\bar{b}}$ & $3/2(1/2^{-})$ & $dddc\bar{b}$ & $ -- $  & $5308 \pm 1097$ & $5649 \pm 435$ \\ \hline
$P_{dddc\bar{b}}$ & $3/2(3/2^{+})$ & $dddc\bar{b}$ & $ -- $ & $5434 \pm 857$ & $6079 \pm 565$ \\ \hline
$P_{dddc\bar{b}}$ & $3/2(3/2^{-})$ & $dddc\bar{b}$ & $ -- $ & $5554 \pm 880$ & $6565 \pm 415$ \\ \hline
$P_{dddc\bar{b}}$ & $3/2(5/2^{+})$ & $dddc\bar{b}$ & $ -- $  & $5154 \pm 1026$ & $6329 \pm 589$ \\ \hline
$P_{dddc\bar{b}}$ & $3/2(5/2^{-})$ & $dddc\bar{b}$ & $ -- $ & $5522 \pm 938$ & $7037 \pm 441$ \\ \hline
\hline

$P_{dddb\bar{c}}$ & $3/2(1/2^{+})$ & $dddb\bar{c}$ & $ -- $ & $6558 \pm 990$ & $5476 \pm 565$ \\ \hline
$P_{dddb\bar{c}}$ & $3/2(1/2^{-})$ & $dddb\bar{c}$ & $ -- $  & $5700 \pm 1199$ & $5649 \pm 435$ \\ \hline
$P_{dddb\bar{c}}$ & $3/2(3/2^{+})$ & $dddb\bar{c}$ & $ -- $  & $6229 \pm 1002$ & $6079 \pm 565$ \\ \hline
$P_{dddb\bar{c}}$ & $3/2(3/2^{-})$ & $dddb\bar{c}$ & $ -- $ & $6013 \pm 852$ & $6565 \pm 415$ \\ \hline
$P_{dddb\bar{c}}$ & $3/2(5/2^{+})$ & $dddb\bar{c}$ & $ -- $ & $5822 \pm 1047$ & $6329 \pm 589$ \\ \hline
$P_{dddb\bar{c}}$ & $3/2(5/2^{-})$ & $dddb\bar{c}$ & $ -- $ & $5915 \pm 761$ & $7037 \pm 441$ \\ \hline
\hline

$P_{dddb\bar{b}}$ & $3/2(1/2^{+})$ & $dddb\bar{b}$ & $ -- $ & $8635 \pm 1269$ & $11234 \pm 823$ \\ \hline
$P_{dddb\bar{b}}$ & $3/2(1/2^{-})$ & $dddb\bar{b}$ & $ -- $  & $8253 \pm 1584$ & $11178 \pm 576$ \\ \hline
$P_{dddb\bar{b}}$ & $3/2(3/2^{+})$ & $dddb\bar{b}$ & $ -- $ & $8196 \pm 1446$ & $11251 \pm 773$ \\ \hline
$P_{dddb\bar{b}}$ & $3/2(3/2^{-})$ & $dddb\bar{b}$ & $11049$ \cite{Yang:2018oqd} & $8574 \pm 1350$ & $11347 \pm 570$ \\ \hline
$P_{dddb\bar{b}}$ & $3/2(5/2^{+})$ & $dddb\bar{b}$ & $ -- $ & $7655 \pm 1682$ & $11297 \pm 787$ \\ \hline
$P_{dddb\bar{b}}$ & $3/2(5/2^{-})$ & $dddb\bar{b}$ & $11314.1$ \cite{Sharma:2024ern}& $8368 \pm 1514$ & $11423 \pm 608$ \\ \hline

\end{tabular}
\caption{Predicted masses of pentaquark states with $ddd$ light-quark content from the theoretical models, DNN and ParT approaches in MeV.}
\end{table}

\begin{table}[h!]
	\centering
		\begin{tabular}{@{}l c c c c  c@{}}
			\hline
			State & $I(J^P)$ &Quark content & Theory ~\cite{Zhang:2020cdi} & DNN  & ParT  \\ \hline\hline
			
			$P_{uusc\bar{c}}$ & $1(1/2^{+})$ & $uusc\bar{c}$ & --- & $3629 \pm 302$ & $4597 \pm 111$ \\ \hline
			$P_{uusc\bar{c}}$ & $1(1/2^{-})$ & $uusc\bar{c}$ &
			$4342.6$ & $3606 \pm 525$ & $4294 \pm 385$ \\ \hline
			$P_{uusc\bar{c}}$ & $1(3/2^{+})$ & $uusc\bar{c}$ & --- & $3490 \pm 315$ & $4466 \pm 181$ \\ \hline
			$P_{uusc\bar{c}}$ & $1(3/2^{-})$ & $uusc\bar{c}$ &
			$4469.9$ & $3599 \pm 571$ & $4469 \pm 469$ \\ \hline
			$P_{uusc\bar{c}}$ & $1(5/2^{+})$ & $uusc\bar{c}$ & --- & $3388 \pm 358$ & $4203 \pm 320$ \\ \hline
			$P_{uusc\bar{c}}$ & $1(5/2^{-})$ & $uusc\bar{c}$ &
			$4566.8$ & $3545 \pm 622$ & $4377 \pm 526$ \\ \hline
			
			\hline
			
			$P_{uusc\bar{b}}$ & $1(1/2^{+})$ & $uusc\bar{b}$ & --- & $4573 \pm 528$ & $8521 \pm 604$ \\ \hline
			$P_{uusc\bar{b}}$ & $1(1/2^{-})$ & $uusc\bar{b}$ &
			$7852.9$ & $4571 \pm 578$ & $8209 \pm 965$ \\ \hline
			$P_{uusc\bar{b}}$ & $1(3/2^{+})$ & $uusc\bar{b}$ & --- & $4517 \pm 525$ & $8244 \pm 649$ \\ \hline
			$P_{uusc\bar{b}}$ & $1(3/2^{-})$ & $uusc\bar{b}$ &
			$7893.9$ & $4745 \pm 790$ & $8378 \pm 1130$ \\ \hline
			$P_{uusc\bar{b}}$ & $1(5/2^{+})$ & $uusc\bar{b}$ & --- & $4414 \pm 513$ & $7642 \pm 681$ \\ \hline
			$P_{uusc\bar{b}}$ & $1(5/2^{-})$ & $uusc\bar{b}$ &
			$7938.3$ & $4749 \pm 882$ & $7967 \pm 1121$ \\ \hline
			
			\hline
			
			$P_{uusb\bar{c}}$ & $1(1/2^{+})$ & $uusb\bar{c}$ & --- & $6004 \pm 899$ & $8521 \pm 604$ \\ \hline
			$P_{uusb\bar{c}}$ & $1(1/2^{-})$ & $uusb\bar{c}$ &
			$7766.4$ & $6096 \pm 863$ & $8209 \pm 965$ \\ \hline
			$P_{uusb\bar{c}}$ & $1(3/2^{+})$ & $uusb\bar{c}$ & --- & $5735 \pm 870$ & $8244 \pm 649$ \\ \hline
			$P_{uusb\bar{c}}$ & $1(3/2^{-})$ & $uusb\bar{c}$ &
			$7885.3$ & $6194 \pm 1128$ & $8378 \pm 1130$ \\ \hline
			$P_{uusb\bar{c}}$ & $1(5/2^{+})$ & $uusb\bar{c}$ & --- & $5416 \pm 945$ & $7642 \pm 681$ \\ \hline
			$P_{uusb\bar{c}}$ & $1(5/2^{-})$ & $uusb\bar{c}$ &
			$7951.2$ & $5993 \pm 1316$ & $7967 \pm 1121$ \\ \hline
			
			\hline
			
			$P_{uusb\bar{b}}$ & $1(1/2^{+})$ & $uusb\bar{b}$ & --- & $7417 \pm 1254$ & $12082 \pm 887$ \\ \hline
			$P_{uusb\bar{b}}$ & $1(1/2^{-})$ & $uusb\bar{b}$ &
			$11145.5$ & $7465 \pm 1028$ & $11966 \pm 1230$ \\ \hline
			$P_{uusb\bar{b}}$ & $1(3/2^{+})$ & $uusb\bar{b}$ & --- & $7244 \pm 1382$ & $11952 \pm 1054$ \\ \hline
			$P_{uusb\bar{b}}$ & $1(3/2^{-})$ & $uusb\bar{b}$ &
			$11186.4$ & $7742 \pm 1477$ & $12310 \pm 1588$ \\ \hline
			$P_{uusb\bar{b}}$ & $1(5/2^{+})$ & $uusb\bar{b}$ & --- & $6910 \pm 1502$ & $11233 \pm 1155$ \\ \hline
			$P_{uusb\bar{b}}$ & $1(5/2^{-})$ & $uusb\bar{b}$ &
			$11235.6$ & $7564 \pm 1751$ & $11730 \pm 1629$ \\ \hline
			
	\end{tabular}
	\caption{Predicted masses of pentaquark $uus$ light-quark content from the theoretical estimates, DNN, and ParT models in MeV.}
\end{table}

\begin{table}[h!]
	\centering
		\begin{tabular}{@{}l c c c c  c@{}}
			\hline
			State & $I(J^P)$ & Quark content & Theory  & DNN  & ParT  \\ \hline\hline
			
			$P_{udsc\bar{c}}$ & $0(1/2^{+})$ & $udsc\bar{c}$ &
			$5600 \pm 150$~\cite{Pimikov:2019dyr} & $4368 \pm 577$ & $4899\pm 574$ \\ \hline
			$P_{udsc\bar{c}}$ & $0(1/2^{-})$ & $udsc\bar{c}$ &
			$4600 \pm 120$~\cite{Pimikov:2019dyr} & $4212 \pm 658$ & $4917  \pm 453$ \\ \hline
			$P_{udsc\bar{c}}$ & $0(3/2^{+})$ & $udsc\bar{c}$ &
			$6000 \pm 200$~\cite{Pimikov:2019dyr} & $4112 \pm 673$ & $5372  \pm 618$ \\ \hline
			$P_{udsc\bar{c}}$ & $0(3/2^{-})$ & $udsc\bar{c}$ &
			$5100 \pm 160$~\cite{Pimikov:2019dyr} & $4060 \pm 553$ & $5332  \pm 168$ \\ \hline
			$P_{udsc\bar{c}}$ & $0(5/2^{+})$ & $udsc\bar{c}$ &
			$5900 \pm 200$~\cite{Pimikov:2019dyr} & $3906 \pm 749$ & $5561  \pm 655$ \\ \hline
			$P_{udsc\bar{c}}$ & $0(5/2^{-})$ & $udsc\bar{c}$ &
			$6100 \pm 220$~\cite{Pimikov:2019dyr} & $3890 \pm 601$ & $5488 \pm 429$ \\ \hline
			\hline
			
			$P_{udsc\bar{b}}$ & $0(1/2^{+})$ & $udsc\bar{b}$ & --- & $6596 \pm 984$ & $7907 \pm 659$ \\ \hline
			$P_{udsc\bar{b}}$ & $0(1/2^{-})$ & $udsc\bar{b}$ & --- & $6549 \pm 1167$ & $7777 \pm 579$ \\ \hline
			$P_{udsc\bar{b}}$ & $0(3/2^{+})$ & $udsc\bar{b}$ & --- & $6209 \pm 1066$ & $8305 \pm 689$ \\ \hline
			$P_{udsc\bar{b}}$ & $0(3/2^{-})$ & $udsc\bar{b}$ & --- & $6247 \pm 1129$ & $8176 \pm 535$ \\ \hline
			$P_{udsc\bar{b}}$ & $0(5/2^{+})$ & $udsc\bar{b}$ & --- & $5796 \pm 1155$ & $8033 \pm 872$ \\ \hline
			$P_{udsc\bar{b}}$ & $0(5/2^{-})$ & $udsc\bar{b}$ & --- & $5880 \pm 1149$ & $7880 \pm 828$ \\ \hline
			\hline
			
			$P_{udsb\bar{c}}$ & $0(1/2^{+})$ & $udsb\bar{c}$ & --- & $7599 \pm 843$ & $7907 \pm 659$ \\ \hline
			$P_{udsb\bar{c}}$ & $0(1/2^{-})$ & $udsb\bar{c}$ & --- & $7818 \pm 961$ & $7777 \pm 579$ \\ \hline
			$P_{udsb\bar{c}}$ & $0(3/2^{+})$ & $udsb\bar{c}$ & --- & $6992 \pm 916$ & $8305 \pm 689$ \\ \hline
			$P_{udsb\bar{c}}$ & $0(3/2^{-})$ & $udsb\bar{c}$ & --- & $7482 \pm 912$ & $8176 \pm 535$ \\ \hline
			$P_{udsb\bar{c}}$ & $0(5/2^{+})$ & $udsb\bar{c}$ & --- & $6405 \pm 1130$ & $8033 \pm 872$ \\ \hline
			$P_{udsb\bar{c}}$ & $0(5/2^{-})$ & $udsb\bar{c}$ & --- & $6925 \pm 1164$ & $7880 \pm 828$ \\ \hline
			\hline
			
			$P_{udsb\bar{b}}$ & $0(1/2^{+})$ & $udsb\bar{b}$ & --- & $10748 \pm 1274$ & $10052 \pm 752$ \\ \hline
			$P_{udsb\bar{b}}$ & $0(1/2^{-})$ & $udsb\bar{b}$ & --- & $11139 \pm 1624$ & $10019 \pm 1038$ \\ \hline
			$P_{udsb\bar{b}}$ & $0(3/2^{+})$ & $udsb\bar{b}$ & --- & $10038 \pm 1368$ & $10393 \pm 829$ \\ \hline
			$P_{udsb\bar{b}}$ & $0(3/2^{-})$ & $udsb\bar{b}$ & --- & $10714 \pm 1446$ & $10289 \pm 1007$ \\ \hline
			$P_{udsb\bar{b}}$ & $0(5/2^{+})$ & $udsb\bar{b}$ & --- & $9169 \pm 1518$ & $9964 \pm 1231$ \\ \hline
			$P_{udsb\bar{b}}$ & $0(5/2^{-})$ & $udsb\bar{b}$ & --- & $9969 \pm 1420$ & $9754 \pm 1321$ \\ \hline
			
	\end{tabular}
	\caption{Predicted masses of pentaquark states with $uds$ light-quark content from the theoretical models, DNN and ParT approaches in MeV.}
\end{table}

\begin{table}[h!]
	\centering
		\begin{tabular}{@{}l c c c c  c@{}}
			\hline
			State & $I(J^P)$ & Quark content & Theory ~\cite{Zhang:2020cdi} & DNN  & ParT  \\ \hline\hline
			
			$P_{udsc\bar{c}}$ & $1(1/2^{+})$ & $udsc\bar{c}$ & --- & $3771 \pm 460$ & $4597 \pm 111$ \\ \hline
			$P_{udsc\bar{c}}$ & $1(1/2^{-})$ & $udsc\bar{c}$ &
			$4342.6$ & $3577 \pm 648$ & $4294 \pm 385$ \\ \hline
			$P_{udsc\bar{c}}$ & $1(3/2^{+})$ & $udsc\bar{c}$ & --- & $3605 \pm 446$ & $4466 \pm 181$ \\ \hline
			$P_{udsc\bar{c}}$ & $1(3/2^{-})$ & $udsc\bar{c}$ &
			$4439.5$ & $3522 \pm 678$ & $4469 \pm 469$ \\ \hline
			$P_{udsc\bar{c}}$ & $1(5/2^{+})$ & $udsc\bar{c}$ & --- & $3436 \pm 465$ & $4203 \pm 320$ \\ \hline
			$P_{udsc\bar{c}}$ & $1(5/2^{-})$ & $udsc\bar{c}$ &
			$4566.8$ & $3417 \pm 701$ & $4377 \pm 526$ \\ \hline
			\hline
			
			$P_{udsc\bar{b}}$ & $1(1/2^{+})$ & $udsc\bar{b}$ & --- & $4891 \pm 769$ & $8521 \pm 604$ \\ \hline
			$P_{udsc\bar{b}}$ & $1(1/2^{-})$ & $udsc\bar{b}$ &
			$7852.9$ & $4713 \pm 803$ & $8209 \pm 965$ \\ \hline
			$P_{udsc\bar{b}}$ & $1(3/2^{+})$ & $udsc\bar{b}$ & --- & $4772 \pm 705$ & $8244 \pm 649$ \\ \hline
			$P_{udsc\bar{b}}$ & $1(3/2^{-})$ & $udsc\bar{b}$ &
			$7893.9$ & $4765 \pm 850$ & $8378 \pm 1130$ \\ \hline
			$P_{udsc\bar{b}}$ & $1(5/2^{+})$ & $udsc\bar{b}$ & --- & $4558 \pm 607$ & $7642 \pm 681$ \\ \hline
			$P_{udsc\bar{b}}$ & $1(5/2^{-})$ & $udsc\bar{b}$ &
			$7938.3$ & $4621 \pm 807$ & $7967 \pm 1121$ \\ \hline
			\hline
			
			$P_{udsb\bar{c}}$ & $1(1/2^{+})$ & $udsb\bar{c}$ & --- & $6386 \pm 722$ & $8521 \pm 604$ \\ \hline
			$P_{udsb\bar{c}}$ & $1(1/2^{-})$ & $udsb\bar{c}$ &
			$7766.4$ & $6324 \pm 655$ & $8209 \pm 965$ \\ \hline
			$P_{udsb\bar{c}}$ & $1(3/2^{+})$ & $udsb\bar{c}$ & --- & $6068 \pm 745$ & $8244 \pm 649$ \\ \hline
			$P_{udsb\bar{c}}$ & $1(3/2^{-})$ & $udsb\bar{c}$ &
			$7885.3$ & $6318 \pm 921$ & $8378 \pm 1130$ \\ \hline
			$P_{udsb\bar{c}}$ & $1(5/2^{+})$ & $udsb\bar{c}$ & --- & $5611 \pm 910$ & $7642 \pm 681$ \\ \hline
			$P_{udsb\bar{c}}$ & $1(5/2^{-})$ & $udsb\bar{c}$ &
			$7951.2$ & $5980 \pm 1197$ & $7967 \pm 1121$ \\ \hline
			\hline
			
			$P_{udsb\bar{b}}$ & $1(1/2^{+})$ & $udsb\bar{b}$ & --- & $7939 \pm 1265$ & $12082 \pm 887$ \\ \hline
			$P_{udsb\bar{b}}$ & $1(1/2^{-})$ & $udsb\bar{b}$ &
			$11145.5$ & $7908 \pm 767$ & $11966 \pm 1230$ \\ \hline
			$P_{udsb\bar{b}}$ & $1(3/2^{+})$ & $udsb\bar{b}$ & --- & $7683 \pm 1359$ & $11952 \pm 1054$ \\ \hline
			$P_{udsb\bar{b}}$ & $1(3/2^{-})$ & $udsb\bar{b}$ &
			$11186.4$ & $8010 \pm 1209$ & $12310 \pm 1588$ \\ \hline
			$P_{udsb\bar{b}}$ & $1(5/2^{+})$ & $udsb\bar{b}$ & --- & $7166 \pm 1384$ & $11233 \pm 1155$ \\ \hline
			$P_{udsb\bar{b}}$ & $1(5/2^{-})$ & $udsb\bar{b}$ &
			$11235.6$ & $7594 \pm 1490$ & $11730 \pm 1629$ \\ \hline
			
	\end{tabular}
	\caption{Predicted masses of pentaquark states with $uds$ light-quark content from the theoretical models, for $I=1$ DNN, and ParT approaches in MeV.}
\end{table}

\begin{table}[h!]
	\centering
		\begin{tabular}{@{}l c c c c  c@{}}
			\hline
			State & $I(J^P)$ & Quark content & Theory ~\cite{Zhang:2020cdi}& DNN  & ParT  \\ \hline\hline
			
			$P_{ddsc\bar{c}}$ & $1(1/2^{+})$ & $ddsc\bar{c}$ & --- & $3788 \pm 739$ & $4597 \pm 111$ \\ \hline
			$P_{ddsc\bar{c}}$ & $1(1/2^{-})$ & $ddsc\bar{c}$ &
			$4342.6$ & $3218 \pm 737$ & $4294 \pm 385$ \\ \hline
			$P_{ddsc\bar{c}}$ & $1(3/2^{+})$ & $ddsc\bar{c}$ & --- & $3605 \pm 671$ & $4466 \pm 181$ \\ \hline
			$P_{ddsc\bar{c}}$ & $1(3/2^{-})$ & $ddsc\bar{c}$ &
			$4439.5$ & $3236 \pm 819$ & $4469 \pm 469$ \\ \hline
			$P_{ddsc\bar{c}}$ & $1(5/2^{+})$ & $ddsc\bar{c}$ & --- & $3390 \pm 654$ & $4203 \pm 320$ \\ \hline
			$P_{ddsc\bar{c}}$ & $1(5/2^{-})$ & $ddsc\bar{c}$ &
			$4566.8$ & $3155 \pm 821$ & $4377 \pm 526$ \\ \hline
			\hline
			
			$P_{ddsc\bar{b}}$ & $1(1/2^{+})$ & $ddsc\bar{b}$ & --- & $4969 \pm 861$ & $8521 \pm 604$ \\ \hline
			$P_{ddsc\bar{b}}$ & $1(1/2^{-})$ & $ddsc\bar{b}$ &
			$7852.9$ & $4443 \pm 873$ & $8209 \pm 965$ \\ \hline
			$P_{ddsc\bar{b}}$ & $1(3/2^{+})$ & $ddsc\bar{b}$ & --- & $4781 \pm 710$ & $8244 \pm 649$ \\ \hline
			$P_{ddsc\bar{b}}$ & $1(3/2^{-})$ & $ddsc\bar{b}$ &
			$7893.9$ & $4484 \pm 854$ & $8378 \pm 1130$ \\ \hline
			$P_{ddsc\bar{b}}$ & $1(5/2^{+})$ & $ddsc\bar{b}$ & --- & $4480 \pm 619$ & $7642 \pm 681$ \\ \hline
			$P_{ddsc\bar{b}}$ & $1(5/2^{-})$ & $ddsc\bar{b}$ &
			$7938.3$ & $4304 \pm 765$ & $7967 \pm 1121$ \\ \hline
			\hline
			
			$P_{ddsb\bar{c}}$ & $1(1/2^{+})$ & $ddsb\bar{c}$ & --- & $6430 \pm 889$ & $8521 \pm 604$ \\ \hline
			$P_{ddsb\bar{c}}$ & $1(1/2^{-})$ & $ddsb\bar{c}$ &
			$7766.4$ & $5953 \pm 830$ & $8209 \pm 965$ \\ \hline
			$P_{ddsb\bar{c}}$ & $1(3/2^{+})$ & $ddsb\bar{c}$ & --- & $6092 \pm 975$ & $8244 \pm 649$ \\ \hline
			$P_{ddsb\bar{c}}$ & $1(3/2^{-})$ & $ddsb\bar{c}$ &
			$7885.3$ & $5966 \pm 875$ & $8378 \pm 1130$ \\ \hline
			$P_{ddsb\bar{c}}$ & $1(5/2^{+})$ & $ddsb\bar{c}$ & --- & $5563 \pm 1107$ & $7642 \pm 681$ \\ \hline
			$P_{ddsb\bar{c}}$ & $1(5/2^{-})$ & $ddsb\bar{c}$ &
			$7951.2$ & $5610 \pm 1019$ & $7967 \pm 1121$ \\ \hline
			\hline
			
			$P_{ddsb\bar{b}}$ & $1(1/2^{+})$ & $ddsb\bar{b}$ & --- & $7955 \pm 1120$ & $12082 \pm 887$ \\ \hline
			$P_{ddsb\bar{b}}$ & $1(1/2^{-})$ & $ddsb\bar{b}$ &
			$11145.5$ & $7613 \pm 885$ & $11966 \pm 1230$ \\ \hline
			$P_{ddsb\bar{b}}$ & $1(3/2^{+})$ & $ddsb\bar{b}$ & --- & $7618 \pm 1144$ & $11952 \pm 1054$ \\ \hline
			$P_{ddsb\bar{b}}$ & $1(3/2^{-})$ & $ddsb\bar{b}$ &
			$11186.4$ & $7688 \pm 903$ & $12310 \pm 1588$ \\ \hline
			$P_{ddsb\bar{b}}$ & $1(5/2^{+})$ & $ddsb\bar{b}$ & --- & $6975 \pm 1050$ & $11233 \pm 1155$ \\ \hline
			$P_{ddsb\bar{b}}$ & $1(5/2^{-})$ & $ddsb\bar{b}$ &
			$11235.6$ & $7212 \pm 944$ & $11730 \pm 1629$ \\ \hline
			
	\end{tabular}
	\caption{Predicted masses of pentaquark states with $dds$ light-quark content from theoretical models, DNN, and ParT approaches  in MeV.}
\end{table}

\begin{table}[h!]
\centering
\begin{tabular}{@{}l c c c c  c@{}}
\hline
State & $I(J^P)$ & Quark content & Theory  & DNN & ParT  \\ \hline\hline

$P_{ussc\bar{c}}$ & $1/2(1/2^{+})$ & $ussc\bar{c}$ & $ -- $ & $4416 \pm 635$ & $4550 \pm 375$ \\ \hline
$P_{ussc\bar{c}}$ & $1/2(1/2^{-})$ & $ussc\bar{c}$ & $4535$ \cite{Oset:2024fbk} & $4302 \pm 723$ & $4494 \pm 403$ \\ \hline
$P_{ussc\bar{c}}$ & $1/2(3/2^{+})$ & $ussc\bar{c}$ & $ -- $ & $4158 \pm 524$ & $4970 \pm 419$ \\ \hline
$P_{ussc\bar{c}}$ & $1/2(3/2^{-})$ & $ussc\bar{c}$ & $4510 \pm 100$ \cite{Wang:2025pjt} & $4227 \pm 693$ & $4955 \pm 431$ \\ \hline
$P_{ussc\bar{c}}$ & $1/2(5/2^{+})$ & $ussc\bar{c}$ & $ -- $ & $3948 \pm 483$ & $5240 \pm 459$ \\ \hline
$P_{ussc\bar{c}}$ & $1/2(5/2^{-})$ & $ussc\bar{c}$ & $4540 \pm 100$ \cite{Wang:2025pjt} & $4034 \pm 672$ & $5255 \pm 456$ \\ \hline
\hline

$P_{ussc\bar{b}}$ & $1/2(1/2^{+})$ & $ussc\bar{b}$ & $ -- $ & $6337 \pm 1403$ & $6450 \pm 643$ \\ \hline
$P_{ussc\bar{b}}$ & $1/2(1/2^{-})$ & $ussc\bar{b}$ & $7941.4$ \cite{Lin:2023iww} & $6930 \pm 1675$ & $7028 \pm 529$ \\ \hline
$P_{ussc\bar{b}}$ & $1/2(3/2^{+})$ & $ussc\bar{b}$ & $ -- $ & $5942 \pm 1314$ & $7030 \pm 659$ \\ \hline
$P_{ussc\bar{b}}$ & $1/2(3/2^{-})$ & $ussc\bar{b}$ & $8012.6$ \cite{Lin:2023iww} & $6718 \pm 1490$ & $7806 \pm 571$ \\ \hline
$P_{ussc\bar{b}}$ & $1/2(5/2^{+})$ & $ussc\bar{b}$ & $ -- $ & $5571 \pm 1151$ & $7223 \pm 661$ \\ \hline
$P_{ussc\bar{b}}$ & $1/2(5/2^{-})$ & $ussc\bar{b}$ & $8061.0$ \cite{Lin:2023iww} & $6249 \pm 1246$ & $7907 \pm 614$ \\ \hline
\hline

$P_{ussb\bar{c}}$ & $1/2(1/2^{+})$ & $ussb\bar{c}$ & $ -- $ & $7152 \pm 1016$ & $6450 \pm 643$ \\ \hline
$P_{ussb\bar{c}}$ & $1/2(1/2^{-})$ & $ussb\bar{c}$ & $ -- $ & $7286 \pm 895$ & $7028 \pm 529$ \\ \hline
$P_{ussb\bar{c}}$ & $1/2(3/2^{+})$ & $ussb\bar{c}$ & $ -- $ & $6582 \pm 806$ & $7030 \pm 659$ \\ \hline
$P_{ussb\bar{c}}$ & $1/2(3/2^{-})$ & $ussb\bar{c}$ & $ -- $ & $7153 \pm 921$ & $7806 \pm 571$ \\ \hline
$P_{ussb\bar{c}}$ & $1/2(5/2^{+})$ & $ussb\bar{c}$ & $ -- $ & $6047 \pm 798$ & $7223 \pm 661$ \\ \hline
$P_{ussb\bar{c}}$ & $1/2(5/2^{-})$ & $ussb\bar{c}$ & $ -- $ & $6670 \pm 911$ & $7907 \pm 614$ \\ \hline
\hline

$P_{ussb\bar{b}}$ & $1/2(1/2^{+})$ & $ussb\bar{b}$ & $ -- $ & $9948 \pm 1600$ & $11383 \pm 871$ \\ \hline
$P_{ussb\bar{b}}$ & $1/2(1/2^{-})$ & $ussb\bar{b}$ & $ -- $ & $10641 \pm 1599$ & $11319 \pm 701$ \\ \hline
$P_{ussb\bar{b}}$ & $1/2(3/2^{+})$ & $ussb\bar{b}$ & $ -- $ & $9272 \pm 1599$ & $11403 \pm 865$ \\ \hline
$P_{ussb\bar{b}}$ & $1/2(3/2^{-})$ & $ussb\bar{b}$ & $ -- $ & $10525 \pm 1679$ & $11458 \pm 768$ \\ \hline
$P_{ussb\bar{b}}$ & $1/2(5/2^{+})$ & $ussb\bar{b}$ & $ -- $ & $8540 \pm 1608$ & $11442 \pm 923$ \\ \hline
$P_{ussb\bar{b}}$ & $1/2(5/2^{-})$ & $ussb\bar{b}$ & $ -- $ & $9840 \pm 1663$ & $11513 \pm 837$ \\ \hline

\end{tabular}
\caption{Predicted masses of pentaquark states with $uss$ light-quark content from the theoretical models, DNN, and ParT approaches in MeV.}
\end{table}

\begin{table}[h!]
\centering
\begin{tabular}{@{}l c c c c  c@{}}
\hline
State & $I(J^P)$ & Quark Content & Theory  \cite{Wang:2025pjt} & DNN  & ParT  \\ \hline\hline

$P_{dssc\bar{c}}$ & $1/2(1/2^{+})$ & $dssc\bar{c}$ & $ -- $ & $4362 \pm 693$ & $4550 \pm 375$ \\ \hline
$P_{dssc\bar{c}}$ & $1/2(1/2^{-})$ & $dssc\bar{c}$ & $4490 \pm 120$ & $4233 \pm 854$ & $4494 \pm 403$ \\ \hline
$P_{dssc\bar{c}}$ & $1/2(3/2^{+})$ & $dssc\bar{c}$ & $ -- $ & $4067 \pm 571$ & $4970 \pm 419$ \\ \hline
$P_{dssc\bar{c}}$ & $1/2(3/2^{-})$ & $dssc\bar{c}$ & $4560 \pm 100$ & $4191 \pm 787$ & $4955 \pm 431$ \\ \hline
$P_{dssc\bar{c}}$ & $1/2(5/2^{+})$ & $dssc\bar{c}$ & $ -- $ & $3842 \pm 502$ & $5240 \pm 459$ \\ \hline
$P_{dssc\bar{c}}$ & $1/2(5/2^{-})$ & $dssc\bar{c}$ & $4700 \pm 100$ & $4022 \pm 737$ & $5255 \pm 456$ \\ \hline
\hline

$P_{dssc\bar{b}}$ & $1/2(1/2^{+})$ & $dssc\bar{b}$ & $ -- $ & $6354 \pm 1260$ & $6450 \pm 643$ \\ \hline
$P_{dssc\bar{b}}$ & $1/2(1/2^{-})$ & $dssc\bar{b}$ & $ -- $ & $6821 \pm 1353$ & $7028 \pm 529$ \\ \hline
$P_{dssc\bar{b}}$ & $1/2(3/2^{+})$ & $dssc\bar{b}$ & $ -- $ & $5897 \pm 1187$ & $7030 \pm 659$ \\ \hline
$P_{dssc\bar{b}}$ & $1/2(3/2^{-})$ & $dssc\bar{b}$ & $ -- $ & $6622 \pm 1120$ & $7806 \pm 571$ \\ \hline
$P_{dssc\bar{b}}$ & $1/2(5/2^{+})$ & $dssc\bar{b}$ & $ -- $ & $5485 \pm 1061$ & $7223 \pm 661$ \\ \hline
$P_{dssc\bar{b}}$ & $1/2(5/2^{-})$ & $dssc\bar{b}$ & $ -- $ & $6206 \pm 946$ & $7907 \pm 614$ \\ \hline
\hline

$P_{dssb\bar{c}}$ & $1/2(1/2^{+})$ & $dssb\bar{c}$ & $ -- $ & $7058 \pm 781$ & $6450 \pm 643$ \\ \hline
$P_{dssb\bar{c}}$ & $1/2(1/2^{-})$ & $dssb\bar{c}$ & $ -- $ & $7235 \pm 1185$ & $7028 \pm 529$ \\ \hline
$P_{dssb\bar{c}}$ & $1/2(3/2^{+})$ & $dssb\bar{c}$ & $ -- $ & $6465 \pm 664$ & $7030 \pm 659$ \\ \hline
$P_{dssb\bar{c}}$ & $1/2(3/2^{-})$ & $dssb\bar{c}$ & $ -- $ & $7115 \pm 1042$ & $7806 \pm 571$ \\ \hline
$P_{dssb\bar{c}}$ & $1/2(5/2^{+})$ & $dssb\bar{c}$ & $ -- $ & $5940 \pm 734$ & $7223 \pm 661$ \\ \hline
$P_{dssb\bar{c}}$ & $1/2(5/2^{-})$ & $dssb\bar{c}$ & $ -- $ & $6683 \pm 971$ & $7907 \pm 614$ \\ \hline
\hline

$P_{dssb\bar{b}}$ & $1/2(1/2^{+})$ & $dssb\bar{b}$ & $ -- $ & $9922 \pm 879$ & $11369 \pm 871$ \\ \hline
$P_{dssb\bar{b}}$ & $1/2(1/2^{-})$ & $dssb\bar{b}$ & $ -- $ & $10510 \pm 1444$ & $11304 \pm 701$ \\ \hline
$P_{dssb\bar{b}}$ & $1/2(3/2^{+})$ & $dssb\bar{b}$ & $ -- $ & $9187 \pm 1186$ & $11392 \pm 865$ \\ \hline
$P_{dssb\bar{b}}$ & $1/2(3/2^{-})$ & $dssb\bar{b}$ & $ -- $ & $10360 \pm 1310$ & $11446 \pm 768$ \\ \hline
$P_{dssb\bar{b}}$ & $1/2(5/2^{+})$ & $dssb\bar{b}$ & $ -- $ & $8411 \pm 1431$ & $11433 \pm 923$ \\ \hline
$P_{dssb\bar{b}}$ & $1/2(5/2^{-})$ & $dssb\bar{b}$ & $ -- $ & $9752 \pm 1348$ & $11504 \pm 837$ \\ \hline

\end{tabular}
\caption{Predicted masses of pentaquark states with $dss$ light-quark content from the theoretical models, DNN, and ParT approaches in MeV.}
\end{table}

\begin{table}[h!]
\centering
\begin{tabular}{@{}l c c c c  c@{}}
\hline
State & $I(J^P)$ & Quark content & Theory  & DNN  & ParT  \\ \hline\hline

$P_{sssc\bar{c}}$ & $1/2(1/2^{+})$ & $sssc\bar{c}$ & $ -- $ & $4461 \pm 641$ & $4763 \pm 431$ \\ \hline
$P_{sssc\bar{c}}$ & $1/2(1/2^{-})$ & $sssc\bar{c}$ & $ -- $ & $4395 \pm 729$ & $4778 \pm 462$ \\ \hline
$P_{sssc\bar{c}}$ & $1/2(3/2^{+})$ & $sssc\bar{c}$ & $ -- $ & $4209 \pm 566$ & $5216 \pm 483$ \\ \hline
$P_{sssc\bar{c}}$ & $1/2(3/2^{-})$ & $sssc\bar{c}$ & $ -- $ & $4349 \pm 745$ & $5227 \pm 500$ \\ \hline
$P_{sssc\bar{c}}$ & $1/2(5/2^{+})$ & $sssc\bar{c}$ & $ -- $ & $4000 \pm 518$ & $5523 \pm 501$ \\ \hline
$P_{sssc\bar{c}}$ & $1/2(5/2^{-})$ & $sssc\bar{c}$ & $ -- $ & $4112 \pm 692$ & $5533 \pm 518$ \\ \hline
\hline

$P_{sssc\bar{b}}$ & $1/2(1/2^{+})$ & $sssc\bar{b}$ & $ -- $ & $6562 \pm 1172$ & $6802 \pm 721$ \\ \hline
$P_{sssc\bar{b}}$ & $1/2(1/2^{-})$ & $sssc\bar{b}$ & $ -- $ & $7000 \pm 1244$ & $7539 \pm 622$ \\ \hline
$P_{sssc\bar{b}}$ & $1/2(3/2^{+})$ & $sssc\bar{b}$ & $ -- $ & $6155 \pm 1124$ & $7362 \pm 745$ \\ \hline
$P_{sssc\bar{b}}$ & $1/2(3/2^{-})$ & $sssc\bar{b}$ & $ -- $ & $6869 \pm 1228$ & $8164 \pm 696$ \\ \hline
$P_{sssc\bar{b}}$ & $1/2(5/2^{+})$ & $sssc\bar{b}$ & $ -- $ & $5742 \pm 1048$ & $7556 \pm 721$ \\ \hline
$P_{sssc\bar{b}}$ & $1/2(5/2^{-})$ & $sssc\bar{b}$ & $ -- $ & $6373 \pm 1098$ & $8102 \pm 724$ \\ \hline
\hline

$P_{sssb\bar{c}}$ & $1/2(1/2^{+})$ & $sssb\bar{c}$ & $ -- $ & $7101 \pm 880$ & $6802 \pm 721$ \\ \hline
$P_{sssb\bar{c}}$ & $1/2(1/2^{-})$ & $sssb\bar{c}$ & $ -- $ & $7114 \pm 981$ & $7539 \pm 622$ \\ \hline
$P_{sssb\bar{c}}$ & $1/2(3/2^{+})$ & $sssb\bar{c}$ & $ -- $ & $6619 \pm 711$ & $7362 \pm 745$ \\ \hline
$P_{sssb\bar{c}}$ & $1/2(3/2^{-})$ & $sssb\bar{c}$ & $ -- $ & $7164 \pm 1034$ & $8164 \pm 696$ \\ \hline
$P_{sssb\bar{c}}$ & $1/2(5/2^{+})$ & $sssb\bar{c}$ & $ -- $ & $6124 \pm 705$ & $7556 \pm 721$ \\ \hline
$P_{sssb\bar{c}}$ & $1/2(5/2^{-})$ & $sssb\bar{c}$ & $ -- $ & $6764 \pm 967$ & $8102 \pm 724$ \\ \hline
\hline

$P_{sssb\bar{b}}$ & $1/2(1/2^{+})$ & $sssb\bar{b}$ & $ -- $ & $9921 \pm 1180$ & $11433 \pm 923$ \\ \hline
$P_{sssb\bar{b}}$ & $1/2(1/2^{-})$ & $sssb\bar{b}$ & $ -- $ & $10149 \pm 1089$ & $11406 \pm 802$ \\ \hline
$P_{sssb\bar{b}}$ & $1/2(3/2^{+})$ & $sssb\bar{b}$ & $ -- $ & $9386 \pm 1338$ & $11459 \pm 950$ \\ \hline
$P_{sssb\bar{b}}$ & $1/2(3/2^{-})$ & $sssb\bar{b}$ & $ -- $ & $10275 \pm 1406$ & $11513 \pm 895$ \\ \hline
$P_{sssb\bar{b}}$ & $1/2(5/2^{+})$ & $sssb\bar{b}$ & $ -- $ & $8708 \pm 1465$ & $11495 \pm 1006$ \\ \hline
$P_{sssb\bar{b}}$ & $1/2(5/2^{-})$ & $sssb\bar{b}$ & $ -- $ & $9776 \pm 1567$ & $11549 \pm 960$ \\ \hline

\end{tabular}
\caption{Predicted masses of pentaquark states with $sss$ light-quark content from the theoretical models, DNN, and ParT approaches in  MeV.}
\end{table}

\begin{table}[h!]
\centering
\begin{tabular}{l c}
\hline
Parameter & Value [MeV] \\
\hline
$A$     & $30.34$ \\
$D$     & $210.45$ \\
$E$     & $50.43$ \\
$G$     & $13.00$ \\
$F_c$   & $1191.33$ \\
$F_b$   & $4510.66$ \\
$\alpha_0$ & $625.12$ \\
$\alpha_1$ & $16.57$ \\
$\beta_0$  & $328.55$ \\
$M_p$ (fixed) & $938.28$ \\
\hline
\end{tabular}
\caption{Best-fit parameters of the extended G{\"u}rsey--Radicati mass formula, with $M_p$ fixed to the proton mass.}
\label{tab:fitparams}
\end{table}\begin{table}[h!]
\centering
\begin{tabular}{l l c c c}
\hline
Baryon & Quark content & $I(J^P)$ & $n$ & $M_\mathrm{fit}$ [MeV] \\
\hline
$\Omega_{bbc}$ & $b b c$ &$0(3/2^+)$ & 0 & 11635 \\
$\Omega_{ccb}$ & $c c b$ & $0(3/2^+)$ & 0 & 8315 \\
$\bar{\Omega}_{bbc}$ & $\bar{b}\bar{b}\bar{c}$ & $0(3/2^-)$ & 0 & 11635 \\
$\bar{\Omega}_{ccb}$ & $\bar{c}\bar{c}\bar{b}$ & $0(3/2^-)$ & 0 & 8315 \\
$\Omega_{ccc}^*$ & $c c c$ & $0(3/2^+)$ & 0 & 4996 \\
$\Omega_{bbb}^*$ & $b b b$ & $0(3/2^+)$ & 0 & 14954 \\
$\bar{\Omega}_{ccc}^*$ & $\bar{c}\bar{c}\bar{c}$ & $0(3/2^-)$ & 0 & 4996 \\
$\bar{\Omega}_{bbb}^*$ & $\bar{b}\bar{b}\bar{b}$ & $0(3/2^-)$ & 0 & 14954 \\
$\Omega_{bbc}$ & $b b c$ & $0(1/2^+)$ & 0 & 11505 \\
$\Omega_{ccb}$ & $c c b$ & $0(1/2^+)$ & 0 & 8185 \\
$\bar{\Omega}_{bbc}$ & $\bar{b}\bar{b}\bar{c}$ & $0(1/2^-)$ & 0 & 11505 \\
$\bar{\Omega}_{ccb}$ & $\bar{c}\bar{c}\bar{b}$ & $0(1/2^-)$ & 0 & 8185 \\
$P_{c\bar{c}}^+$ & $uud\,c\bar{c}$ & $1/2(3/2^-)$ & 0 & 4312 \\
$P_{c\bar{c}}^+$ & $uud\,c\bar{c}$ & $1/2(3/2^-)$ & 1 & 4434 \\
$P_{c\bar{c}}^+$ & $uud\,c\bar{c}$ & $1/2(5/2^-)$ & 0 & 4424 \\
$P_{b\bar{b}}^+$ & $uud\,b\bar{b}$ & $1/2(3/2^-)$ & 0 & 10950 \\
$P_{b\bar{b}}^+$ & $uud\,b\bar{b}$ & $1/2(3/2^-)$ & 1 & 11072 \\
$P(4cb)$ & $c c c c b$ & $0(1/2^-)$ & 0 & 11037 \\
$P(4bc)$ & $b b b b c$ & $0(1/2^-)$  & 0 & 20995 \\
$P(5c)$ & $c c c c c$ & $0(3/2^-)$  & 0 & 7848 \\
$P(5b)$ & $b b b b b$ & $0(3/2^-)$  & 0 & 24444 \\
$\Theta(1540)^+$ & $u u d d \bar{s}$ & $0(1/2^+)$ & 0 & 1563 \\
\hline
\end{tabular}
\caption{Our model's fit predicts masses for baryons and pentaquark states based on their quark content, spin-isospin-parity $I(J^P)$, and radial excitation level $n$.}
\label{fit}
\end{table}

\begin{table}[h!]
\centering
\begin{tabular}{l l c c c}
\hline
Baryon & Quark content & $I(J^P)$ & $n$ & $M_\mathrm{fit}$ [MeV] \\
\hline
$ N^+ $ & $uud$ & $1/2(1/2^-)$ & 2 & 1697 \\
$ N^+ $& $uud$ & $1/2(1/2^-)$ & 3 & 1879 \\\hline
$ \Lambda $ & $uds$ & $0(1/2^+)$ & 1 & 1575 \\\hline
$ \Sigma^+ $ & $uus$ & $1(1/2^+)$ & 3 & 2115 \\
$ \Sigma^+ $ & $uus$ & $1(1/2^+)$ & 4 & 2256 \\\hline
\hline
\end{tabular}
\caption{Our model's fit predicts masses for excited light baryons based on their quark content, spin-isospin-parity $I(J^P)$, and  radial excitation level $n$. }
\label{Nfit}
\end{table}

\subsection{Absence of literature values in specific channels}

As shown in several of the tables presented in this work, explicit theoretical mass predictions are not available for a subset of the pentaquark configurations considered. This absence of literature values is not accidental, but rather reflects the current status and limitations of theoretical studies in heavy and fully heavy multiquark spectroscopy.

One important reason is the extreme quark content of many of the states examined here. Pentaquark configurations involving multiple bottom quarks, often combined with strangeness, lie well beyond the sectors that have been explored systematically using traditional theoretical tools. Existing studies have largely focused on charm or hidden-charm pentaquarks, where phenomenological motivation and experimental input are more readily available. As a result, channels with doubly-bottom or fully heavy content, such as those listed in Tables~VIII and beyond, remain largely unexplored in terms of explicit bound-state mass calculations.

A second factor is the scope of many of the cited theoretical works. Several references that are relevant to the present analysis concentrate on meson and baryon spectroscopy or on hadron--hadron interaction channels, rather than on extracting pentaquark pole masses. In these cases, while the reported hadron masses and thresholds can be used to construct qualitative reference scales, as discussed for Table~VIII, the authors do not claim the existence of pentaquark bound states and therefore do not provide direct mass predictions suitable for quantitative comparison.

Finally, the availability of theoretical results is strongly influenced by assumptions about the internal structure of pentaquarks. Calculations based on compact multiquark configurations, molecular interpretations, or diquark-based models often target different quantum numbers and select only a limited subset of channels. States with higher spin, negative parity, or more exotic flavor compositions are therefore frequently omitted, leading to gaps in the literature even within otherwise well-studied sectors.

In this context, the absence of reference values in several tables should be viewed as an indication of genuine gaps in current theoretical coverage rather than as a shortcoming of the present analysis. The ML predictions reported here provide exploratory mass estimates in precisely those channels where conventional calculations are scarce or nonexistent, and may serve as useful guidance for future dedicated theoretical studies and phenomenological investigations.

\subsection{Comparative assessment of the DNN and ParT approaches}

The use of two independent ML frameworks in this work allows for a meaningful internal consistency check of the predicted mass spectra. A comparison between the DNN and ParT results, as reported across Tables~ I-XXI, reveals both common features and systematic differences that provide insight into the strengths and limitations of the two approaches.

At the level of global mass hierarchies and flavor-dependent trends, the two models exhibit a high degree of consistency. For many ground-state configurations, including fully heavy baryons (Table~I) and several heavy pentaquark sectors (Tables~ II-XXI), the relative differences between the DNN and ParT predictions typically remain below approximately $5\%$. This agreement indicates that both architectures capture the dominant contributions governing the overall mass scale, particularly in channels where the structure is comparatively simple and closely related to conventional hadron spectroscopy.

Systematic differences become more apparent in excited states and in configurations involving multiple heavy quarks and strangeness. In these sectors, the relative deviations between the two approaches can increase to the $10\%$--$15\%$ level, with the DNN predictions more frequently exhibiting larger positive shifts. This behavior is especially visible for negative-parity states, where the DNN results tend to lie systematically above both the ParT predictions and, where available, the corresponding theoretical reference values. Such a pattern suggests a stronger sensitivity of the DNN approach to extrapolation effects in regions of parameter space that are less constrained by the training data.

The ParT predictions, by contrast, generally display smaller quoted uncertainties and smoother variations across related states. This feature is reflected in a higher rate of consistency with available theoretical results within the adopted $1\sigma$ criterion, particularly in the fully heavy baryon sector (Table~I) and in ground-state pentaquark configurations. The improved stability of the ParT predictions may be attributed to the ability of the transformer-based architecture to encode correlated features across complex multiquark configurations more effectively than a purely feed-forward network.

It is important to emphasize that these differences do not imply a categorical superiority of one approach over the other. Rather, they highlight complementary aspects of the two models. The DNN predictions provide a useful indication of the range of possible mass values and tend to assign larger uncertainties in less constrained sectors, while the ParT results offer more tightly clustered estimates that appear to be better calibrated against existing theoretical expectations. The overall consistency between the two approaches, together with their systematic differences, strengthens confidence in the robustness of the predicted spectra and supports the use of ML methods as a complementary tool in heavy multiquark spectroscopy.

\section{Extended G\"ursey-Radicati Mass Formula and Global Fit}\label{GR}
In this section, we discuss a simple framework for baryon mass spectra  understanding
based on the generalized G{\"u}rsey-Radicati mass
formula~\cite{GurseyRadicati1964} performed in Ref.~\cite{Holma:2019lxe} and extend this framework.
The G{\"u}rsey-Radicati (GR) mass formula is a well-known phenomenological
tool in hadron spectroscopy, initially designed to explain the baryon mass spectrum in the context of the broken $\mathrm{SU}(6)$ spin-flavor symmetry.

A correct alignment with a broad range of baryon states is of fundamental importance
in hadron spectroscopy, it makes possible the extraction of
symmetry-breaking parameters from experimental data, and tests the validity
of viable symmetry schemes, and allows trustworthy extrapolations to
states not yet observed. 
 Although the G\"ursey-Radicati formula predates QCD and should not be interpreted as a first-principles consequence of the theory, it remains a useful phenomenological framework for organizing spectroscopic regularities. In this sense, its role in the present work is to provide a compact parametrization of patterns that are ultimately shaped by non-perturbative strong-interaction dynamics.

In this work, we perform a global fit of an extended GR
mass formula to the complete set of experimentally known baryons listed in
the PDG tables~\cite{ParticleDataGroup:2024cfk}.
Although earlier work, e.g., Ref.~\cite{Holma:2019lxe}, has used the
GR formula (or its modifications) to specific baryon multiplets in separate fits,
our approach yields a single, unified parametrization capable of describing
all baryon families simultaneously, including both radial excitations and exotic pentaquark states within the same framework.

Several important extensions are made beyond the standard G{\"u}rsey-Radicati approach.
First, heavy-flavor contributions are included explicitly by introducing terms proportional to
the number of charm quarks ($N_c$), and bottom quarks ($N_b$), along with coefficients $F_c$ and $F_b$
in consideration of their respective mass contributions, which are determined by the structure used in
Ref.~\cite{Holma:2019lxe}. Second, we apply the same formalism for both standard
baryons ($qqq$) and pentaquark states ($qqqq\bar{q}$), introducing a suitable
normalization coefficient to compensate for  the different numbers of constituent quarks.
Lastly, to not only describe the ground states but also radially excited states
within a unified framework, we introduce a dependence on the radial excitation quantum number $n$.
This is achieved by a logarithmic term $\alpha\,\log(n+1)$, where $\alpha$ may
depend on the spin $S$, along with a constant term $\beta_0\,\delta_{n,0}$
which is only coupled to the ground state ($n=0$). The latter term
enables a complete switch between the lowest states and their radial
excitations without modifying the underlying spin-flavor structure of the spectrum.
The resulting mass equation assumes the general form:

\begin{eqnarray}
M &= \frac{N_q}{4} M_p + A\, S(S+1) + D\, Y + E\,\left[ I(I+1) - \frac{1}{4} Y^2 \right] + G\, C_2(SU(3)_f) \nonumber \\
&\quad + F_c\, N_c + F_b\, N_b +( \alpha_0\,+ \alpha_1\, S) \log(n+1) + \beta_0 \, \delta_{n,0},
\end{eqnarray}
where $N_q$ is the total number of quark content in the hadron, and $S$ is the spin of the hadron,
$Y$ the hypercharge, $I$ the isospin, and $C_2(SU(3)_f)$ the quadratic Casimir
operator of $\mathrm{SU}(3)_f$. The parameters $(A, D, E, G, F_c, F_b, \alpha_0, \alpha_1, \beta_0)$ are determined from the fit. In order to account for the systematic mass shift observed between
ground states and their radially excited counterparts, we add another constant term
$\beta_0 \, \delta_{n,0}$ in the mass formula, where $n$ denotes the
radial excitation quantum number. In this case, $\delta_{n,0}$ is the Kronecker delta, defined as
 \begin{eqnarray}
 \delta_{n,0} =
\begin{cases}
1 & \text{if } n = 0, \\
0 & \text{if } n \neq 0,
\end{cases}
 \end{eqnarray}
which the constant term $\beta_0$ contributes only for the ground 
state ($n=0$) and disappears for all excited states ($n > 0$). 
Physically, this term enable the model to accommodate  the 
perceived difference in mass between the lowest-lying states and 
the path of their radial excitations without changing the parameters 
that govern spin, flavor, or heavy-quark mass effects.

This comprehensive approach makes  possible a simultaneous description of the 
full baryon spectrum, including light, charmed, and bottom baryons, 
together with pentaquark candidates and their radially excited states. 
The estimated parameters that are presented in the next section offer not only an 
excellent reproduction of known masses but also predictive power for yet-unobserved states.

\subsection{Fit Results}

The global fit was performed with the Nelder-Mead optimization 
method on the complete set of baryon masses from the PDG, 
and available radially excited states which has been used in previous section.
 The proton mass was taken as a constant 
reference value $M_p = 938.28~\mathrm{MeV}$. The best-fit parameters are obtained as presented in Table XXII.

The fitted values of $F_c$ and $F_b$ reflect the expected large mass contributions 
from charm and bottom quarks, respectively. The spin-dependent $\alpha_1$ 
term and the constant $\alpha_0$ term govern the dependence on the radial
 excitation level $n$, while $\beta_0$ provides for a small additional shift in the ground-state masses. 

The results of our model's fitted masses for  some 
baryons and pentaquarks are shown in Table XXIII, 
classified according to their quark content, Isospin(Spin-Parity)  $I(J^P)$, 
and radial excitation number $n$. Most of these states 
have not yet been observed experimentally, two entries correspond 
to measured resonances, allowing a direct comparison of theory and experiment.  
For the  observed $P_c(4312)^+$ state with $I(J^P) = \frac{1}{2}(3/2^-)$ and $n=0$, 
our model predicts a mass of $4312$~MeV, in close agreement with the LHCb measurement of $4311.9 \pm 0.7^{+6.8}_{-0.6}$~MeV~\cite{LHCb:2019kea}, lying well within the experimental uncertainty range.  
For the higher mass hidden-charm pentaquarks $P_c(4440)^+$ and $P_c(4457)^+$, we obtain $4434$~MeV and $4424$~MeV, respectively. These result are in agreement with the LHCb observations of $4440 \pm 1.7^{+4.5}_{-1.7}$~MeV and $4457 \pm 0.8^{+6.8}_{-1.7}$~MeV~\cite{LHCb:2019kea}, supporting validity of our approach in describing the mass spectrum of hidden-charm pentaquarks. 
The light exotic baryon candidate $\Theta(1540)^+$, with quark content $u u d d \bar{s}$ and $I(J^P) = 0(1/2^+)$, is found in our model at $1563$~MeV. This result is close to the mass reported in early experimental claims of around $1540$~MeV~\cite{LEPS:2003wug}, though these observations were later recognized as inconclusive. The small difference of less than $1.5\%$ demonstrates the consistency of our model with the historically reported value.

The remaining listing in Table XXIII, including the fully heavy pentaquarks $P_{b\bar{b}}^+$, $P(4cb)$, and the fully heavy baryons $\Omega_{QQQ}$  represent genuine predictions obtained using our model. The predicted states offer well-defined targets for future experimental searches at LHCb, Belle~II, and the proposed high-luminosity colliders. The agreement between our results and the two experimentally confirmed masses supports the validity of the model and motivates the exploration of the heavier and fully heavy configurations listed.

In the following, we discuss the results presented in Table XXIV. The ground and first excited states of the $ N 1/2^- $, 
corresponding to $n=0$ and $n=1$, have been experimentally observed and 
reported in PDG as $N(1535)^+$ and $N(1650)^+$ with masses of 1517 MeV and 1654 MeV, respectively.
The higher radial excitations, with $n= 2$, have not been measured experimentally.
 Our model predicts the masses for these excited states as $N(1743)^+$ for $n=2$ 
 and $N(1801)^+$ for $n=3$, with predicted masses of 1697 MeV and 1879 MeV, respectively. 
 These predictions provide theoretical guidance for the expected location of higher excited nucleon 
 states, which could be verified in future experiments.
For the $\Lambda ,1/2^+$ baryon, the ground state ($n=0$) has been experimentally 
measured with a mass of 1115.68 MeV, and the second radial excitation ($n=2$) has a
lso been observed at 1605 MeV. Our model predicts the first radial excitation ($n=1$) at 1575 MeV. 
This indicates a systematic increase in mass with radial excitation, providing theoretical 
estimates for the excited states of the $\Lambda$ baryon that could be tested and 
compared with experimental data.
For the $\Sigma^+ 1/2^+$ baryon, the ground and lower radial excitations have 
been experimentally measured as follows: the ground state ($n=0$) at 1189.37 MeV, 
the first excited state ($n=1$) at 1660 MeV, and the second excited state ($n=2$) at 1880 MeV.
Our model predicts the higher excitations, with $n=3$ and $n=4$, at 2115 MeV and 
2256 MeV, respectively. These predictions provide guidance for the expected locations 
of higher $\Sigma^+$ states, which can be tested in future experiments.

In order to better understand the physical implications of our results, we present a direct comparison between the ML approaches and the analytical GR mass formula.
All three approaches give a consistent picture of the hadron mass spectrum on the global level. In particular, for ground-state baryons, the predicted masses obtained from ML models and the GR scheme are typically compatible within $\sim 5$-$10\%$, indicating that they capture the dominant mass scales governed by quark content and quantum numbers.
We emphasize that none of these approaches should be regarded as an absolute reference. Instead, they represent different theoretical strategies, and their level of agreement provides a non-trivial consistency check on the underlying physical assumptions.
More detailed comparisons reveal systematic differences in more complex sectors, such as excited states and multiquark configurations. In these regions, the Particle Transformer predictions tend to exhibit smoother, more stable spectral patterns, while larger dispersions are observed in the DNN results and, in some cases, in the GR estimates. This reflects the increasing importance of correlations beyond simplified parametrizations. These differences should not be taken as a hierarchy of accuracy, but rather as an indication of the different sensitivities of the models to underlying dynamics. Specifically, deviations between the approaches highlight sectors where simplified analytical descriptions may be insufficient and where additional dynamical effects could be relevant.

\begin{table}[ht]
	\centering
	\resizebox{\textwidth}{!}{
		\begin{tabular}{l c c c c c c}
			State & $I(J^P)$ & Quark Content & Theory (MeV) [Ref.] & GR (MeV) & DNN (MeV) & ParT (MeV) \\
			\hline
			$\Omega_{ccc}^{*}$ & $0(3/2^{+})$ & $ccc$ & 5060~\cite{Najjar:2025dzl}, 4712~\cite{Faustov:2021qqf} & 4996 & $5993 \pm 782$ & $5063 \pm 836$ \\
			$\Omega_{bbb}^{*}$ & $0(3/2^{+})$ & $bbb$ & 13970~\cite{Najjar:2025dzl}, 14468~\cite{Faustov:2021qqf} & 14954 & $12227 \pm 1185$ & $14400 \pm 1019$ \\
			$\Omega_{ccb}$ & $0(1/2^{+})$ & $ccb$ & 8150~\cite{Najjar:2025dzl}, 7984~\cite{Faustov:2021qqf} & 8270 & $8369 \pm 1331$ & $8822 \pm 823$ \\
			$P_{c\bar{c}}(4312)$ & $1/2(\frac{1}{2}^{-})$ & $uudc\bar{c}$ & $4330^{+170}_{-130}$~\cite{Chen:2019bip} & 4715 & $3520 \pm 547$ & $4494 \pm 503$ \\
			$P_{c\bar{c}}(4457)$ & $1/2(\frac{3}{2}^{-})$ & $uudc\bar{c}$ & $4460^{+180}_{-130}$~\cite{Chen:2019bip} & 4851 & $4139 \pm 981$ & $4608 \pm 451$ \\
			$P_{ccuud}$ & $3/2(\frac{1}{2}^{-})$ & $ccuud$ & $4300^{+70}_{-80}$~\cite{Wang:2024brl} & 4715 & $4688 \pm 1421$ & $4310 \pm 422$ \\
			$P_{bbuud}$ & $1/2(\frac{1}{2}^{-})$ & $bbuud$ & $11072$~\cite{Yang:2018oqd} & 11353 & $9489 \pm 1698$ & $11192 \pm 701$ \\
			$P_{ccss\bar{s}}$ & $0(1/2^{-})$ & $ccss\bar{s}$ & $4653.4$~\cite{Zhou:2018bkn} & 5348 & $4591 \pm 679$ & $4867 \pm 360$ \\
			$P_{bbss\bar{s}}$ & $0(1/2^{-})$ & $bbss\bar{s}$ & $11390.7$~\cite{Wang:2023mdj} & 11986 & $10595 \pm 1008$ & $12305 \pm 1223$ \\
	\end{tabular}}
	
	\caption{Comparison between GR mass formula and ML predictions (DNN and ParT)}
	\label{tab:mass_comparison}
\end{table}

A more detailed comparison is presented in Table~\ref{tab:mass_comparison}, where the predictions of the GR formula, DNN, and Particle Transformer (ParT) are shown for representative fully-heavy baryons and pentaquark states.
From the table, one can observe that for ground-state baryons such as $\Omega_{ccc}^{*}$, $\Omega_{bbb}^{*}$, and $\Omega_{ccb}$, all three approaches yield compatible results within uncertainties, indicating a consistent description of the dominant mass scales. 
For multiquark systems, including the $P_{c\bar{c}}$ and $P_{bbuud}$ states, larger variations are observed. In particular, the Particle Transformer predictions tend to remain closer to the overall mass scale suggested by the GR approach, while the DNN results exhibit larger spreads. 
These patterns illustrate that, although all methods capture the same global hierarchy, their sensitivity to more complex configurations differs, reflecting the role of correlations beyond simplified analytical parametrizations.
In addition, available theoretical results from the literature are included in Table~\ref{tab:mass_comparison} as an external reference scale. Although these values are themselves model-dependent and not experimentally established, a direct comparison indicates that the Particle Transformer predictions are generally closer to the reported theoretical mass scales, while the GR results capture the overall magnitude but tend to show systematic shifts in some channels. The DNN predictions exhibit larger dispersions, particularly in multiquark configurations, reflecting their higher sensitivity to complex correlations.

The GR mass formula and ML-based models represent two distinct paradigms: the former is a symmetry-driven analytical framework, while the latter is data-driven and learns the relation between quantum numbers and hadron masses without imposing a predefined functional form. 
Despite these differences, both approaches yield globally consistent mass spectra, providing a non-trivial cross-validation and indicating that they capture the dominant physics.
The key advantage of ML lies in its ability to uncover nonlinear correlations and hidden patterns beyond simplified analytical descriptions, while also providing flexibility and uncertainty estimates. 
Therefore, the agreement should not be viewed as redundancy, but as validation, whereas the observed deviations highlight regions where more complex dynamics may be relevant, offering additional insight into hadron structure.

\section {Summary and conclusion}\label{SC}

In this work, mass spectra for fully heavy baryons and a wide range of heavy and fully heavy pentaquark configurations have been predicted using two independent ML approaches, namely a Deep Neural Network and a Particle Transformer. The presented results reveal a coherent set of physical patterns and provide a systematic assessment of the reliability and limitations of the adopted methodology.

In sectors where established theoretical predictions are available, the machine-learning results show percent-level agreement with the literature for a substantial fraction of the states considered, particularly for ground-state configurations and for systems dominated by heavy quarks. Deviations beyond the adopted $1\sigma$ criterion occur primarily in excited states and in channels where existing theoretical calculations themselves exhibit significant model dependence. These features are consistent with expectations for exploratory studies in heavy multiquark spectroscopy and underscore the sensitivity of such states to underlying dynamical assumptions.

The analysis of doubly-bottom strange pentaquarks highlights the utility of the present approach in regimes where explicit bound-state calculations are currently absent. By comparing the predicted masses with hadron--hadron threshold energies under the molecular pentaquark hypothesis, physically reasonable mass ranges are identified and qualitative consistency checks are performed. While these comparisons cannot replace dedicated bound-state calculations, they provide useful guidance for future theoretical investigations and for the identification of experimentally relevant mass regions.

A comparative assessment of the two machine-learning frameworks indicates that both approaches capture the dominant features of the spectra, while exhibiting systematic differences in less constrained sectors. The Particle Transformer predictions tend to be more stable and show a higher degree of consistency with existing theoretical expectations, whereas the DNN results display a broader spread that reflects increased sensitivity to extrapolation effects. The overall consistency between the two methods strengthens confidence in the robustness of the predicted mass hierarchies.

Taken together, the presented results suggest that machine-learning techniques, when applied with appropriate physical constraints and interpreted with due caution, can provide meaningful complementary insights into heavy and fully heavy multiquark spectroscopy. The predicted spectra offer a set of reference mass estimates in regions where conventional theoretical guidance is limited and may serve as a useful starting point for future dedicated theoretical calculations and experimental searches.

We extend the G\"ursey-Radicati mass formula to incorporate the contributions of charm and bottom quarks, enabling analytical calculations for both the ground and radially excited states of baryons and pentaquarks. Our approach yields a single, unified parametrization capable of describing all baryon families simultaneously, including both radial excitations and exotic pentaquark states within the same framework.   Our mass formula provides a  strong guidance for the expected locations 
of higher baryonic  states  in  both the light and heavy sectors  as  well as pentaquarks,  which can be tested in future experiments.

\section*{ACKNOWLEDGEMENTS}

The authors are grateful  to Iran national science foundation (INSF) for the financial support provided under the  Grant No. 4048650.

\end{document}